 \documentstyle[11pt,aaspp4,flushrt]{article}

\received{}
\accepted{}

\slugcomment{To appear in The Astrophysical Journal}

\lefthead{Bower et al.}
\righthead{Nucleus of M84}

\begin{document}

\title{The Ionization Source in the Nucleus of M84\altaffilmark{1}}

\author{G. A. Bower\altaffilmark{2}, R. F. Green\altaffilmark{2}, 
A. C. Quillen\altaffilmark{3}, A. Danks\altaffilmark{4},
T. Gull\altaffilmark{5}, J. Hutchings\altaffilmark{6},
C. Joseph\altaffilmark{7},\\
M. E. 
Kaiser\altaffilmark{5,8}, 
D. Weistrop\altaffilmark{9}, B. Woodgate\altaffilmark{5}, E. M. Malumuth\altaffilmark{4}, 
and C. Nelson\altaffilmark{9}}

\altaffiltext{1}{Based on observations with the NASA/ESA {\it Hubble Space Telescope},
obtained
at the Space Telescope Science Institute, which is operated by the
Association of Universities for Research in Astronomy, Inc. (AURA), under
NASA contract NAS5-26555.}

\altaffiltext{2}{Kitt Peak National Observatory, National Optical Astronomy Observatories, P. O. 
Box 26732,
Tucson, AZ 85726; gbower@noao.edu, operated by AURA under cooperative agreement with
the National Science Foundation.}
\altaffiltext{3}{Steward Observatory, University of Arizona, Tucson, AZ 85721}
\altaffiltext{4}{Raytheon ITSS, NASA/Goddard Space Flight Center, Code 681, Greenbelt, MD 20771}
\altaffiltext{5}{NASA/Goddard Space Flight Center, Code 681, Greenbelt, MD 20771}
\altaffiltext{6}{Dominion Astrophysical Observatory, 
National Research Council of Canada,
5071 W. Saanich Road, 
Victoria, BC V8X 4M6, Canada}
\altaffiltext{7}{Dept. of Physics \& Astronomy, Rutgers University,
P. O. Box 849, Piscataway, NJ 08855}
\altaffiltext{8}{Department of Physics \& Astronomy, Johns Hopkins University,
Homewood Campus, Baltimore, MD 21218}
\altaffiltext{9}{Department of Physics, University of Nevada, 4505 Maryland
Parkway, Las Vegas, NV 89154}

\newpage

\begin{abstract}

We have obtained new {\it Hubble Space Telescope} (HST) observations of M84, a nearby
massive elliptical galaxy whose nucleus
contains a $\approx 1.5 \times 10^9 \ M_{\odot}$ dark compact object, which presumably is a 
supermassive black hole. Our Space Telescope Imaging Spectrograph (STIS) spectrum provides the
first clear detection of emission lines in the blue (e.g., [O~II] $\lambda 3727$, H$\beta$, and 
[O~III] $\lambda\lambda
4959, 5007$), which arise from a compact region $\approx 0\farcs28$ across centered on the nucleus.
Our Near Infrared Camera and Multi-Object Spectrometer (NICMOS) images
exhibit the best view through the prominent dust lanes evident at optical 
wavelengths and provide a more accurate correction for the internal extinction. The
relative fluxes of the emission lines we have detected in the blue together with 
those detected in the wavelength range $6295 - 6867$ \AA \ by
Bower et al.~(1998, ApJ, 492, L111) indicate that the gas at the nucleus is photoionized by
a nonstellar process, instead of hot
stars. Stellar absorption features from cool stars at the nucleus are very weak. We update the 
spectral energy distribution
of the nuclear point source and find that although it is roughly flat in most bands, the
optical to UV continuum is very red, similar to the spectral energy distribution of BL Lac. 
Thus, the nuclear
point source seen in high-resolution optical images (Bower et al.~1997, ApJ, 483, L33) is not
a star cluster but is instead a nonstellar source. Assuming isotropic emission from this source,
we estimate that the ratio of bolometric luminosity to Eddington luminosity is $\sim 5 \times
10^{-7}$. However, this could be underestimated if this source is a misaligned BL Lac object,
which is a possibility suggested by the spectral energy distribution and the evidence of 
optical variability we describe.

\end{abstract} 


\newpage   

\section{Introduction}

Observations of the nuclei of several nearby galaxies have revealed very compelling evidence 
for massive dark objects with $\sim 10^6 - 10^9 \ M_{\odot}$ (see review in
Richstone et al.~1998), which are presumably 
supermassive black holes (BHs). Since black holes of this mass range are strongly suspected to be
the central engines powering quasars and AGN, this presents an important opportunity to address 
several related problems.
Determining the
local space density of BHs and the host galaxy properties as a function of BH mass 
should yield important insights into the evolution of quasars as a function of cosmic time and
the processes that occur during galaxy formation that might influence BH (and thus quasar) 
formation.
Also, a few of the local BHs reside in galaxies that host an AGN, presenting the
opportunity to relate BH mass to other AGN properties (such as radio jets, nuclear emission line 
regions,
and the spectral energy distribution of the central source).

M84 (NGC~4374 = 3C~272.1) is one such nearby BH candidate galaxy with a relatively weak AGN, 
as indicated by its FR I radio source (Laing \& Bridle~1987), an 82 pc diameter nuclear 
disk of ionized gas aligned approximately perpendicular to the
radio jet axis (Bower et al.~1997, hereafter Paper~I) whose kinematics indicate the presence of
a central $1.5 \times 10^9 \ M_{\odot}$ dark compact object (Bower et al.~1998, hereafter 
Paper~II), and
a nuclear point source seen in optical images (Paper~I). Stellar light profiles in the central
few hundred pc
of high luminosity elliptical galaxies like M84 are shallow power-law cusps (Faber et al.~1997),
suggesting that the sharp nuclear point source in M84 is a distinct entity from the old stellar 
population. It could represent optical emission related to
the AGN powering the radio source, or it could be a nuclear star cluster. 
Spectroscopy in the far-UV bandpass is
sensitive to the presence of hot stars. For example, Maoz et al.~(1998) detected far-UV
absorption lines from massive stars in a few LINERs with compact UV nuclear sources. However, 
detecting
far-UV emission from M84's nucleus would be difficult since the internal extinction is very high. 
The purpose of this
paper is to determine the nature of the nuclear point source by constraining the ionization
mechanism using high spatial resolution STIS spectra of emission lines in the wavelength ranges of
$2900-5700$ \AA \ and $6295-6867$ \AA.
Our blue spectrum thus will provide an indirect diagnostic of the possible presence
of hot stars by determining if the gas is photoionized by hot stars or a nonstellar continuum.     

Since the nuclear point source is embedded
in dust, we also utilize near-infrared images of M84 obtained with NICMOS by Quillen, Bower, \& 
Stritzinger~(1999),
in an effort to measure and to correct for internal extinction.
Throughout this paper, we adopt a distance to M84 of 17 Mpc 
(Mould et al.~1995). At this distance, 
$1''$ corresponds to 82 pc. The 
Galactic extinction along the line of sight is
$A_B = 0\fm13$
(Burstein \& Heiles 1984).

\section{Observations and Data Calibration}

\subsection{Spectroscopy}

Long-slit spectroscopy of M84's nuclear region was obtained with the STIS
CCD, which has a pixel scale of $0\farcs05$/pixel and spatial resolution $\approx
0\farcs1$
(Kimble et al.~1998; Woodgate et al.~1998),
on 1998 April 10.  
M84's nuclear point source
was acquired in the $52'' \times 0\farcs2$ slit oriented at a position angle (P.A.) of
$130\arcdeg$ on the sky, while the dust lanes are oriented at P.A. $\approx 60\arcdeg$ 
on a scale of $1''$ (Paper~I).
After target
acquisition, we obtained an image through the slit,
to verify the slit position.  
We then obtained spectral observations with
the G430L grating, which has a dispersion of 2.73 \AA/pixel.
This grating was set to cover the wavelength range of 2900~\AA \ to
5700~\AA, which includes several prominent stellar absorption lines
(e.g., Mg I b 5167~\AA, 5172~\AA, 5183~\AA, and Ca II H 3968~\AA
\ and K 3933~\AA). The spectral resolution
of our instrumental configuration was 10.9 \AA \ 
(FWHM), assuming uniform illumination of the slit. The presence of a
nuclear point source in optical continuum images
implies that our spectral resolution at the nucleus was
better than this. Model line spread functions based on pre-flight data
predict that the spectral resolution for this observing mode when observing a
point source should be between 3.5~\AA \ at 3500~\AA \ and 4.0~\AA \ at
5500~\AA \ (Walborn \& Baum~1998). 
We integrated for two HST orbits, which was equivalent to $4120$ sec. 
Before the exposure in the second HST orbit, the nucleus was shifted by 4.5 pixels ($0\farcs225$) 
along the slit to allow 
for better rejection of hot pixels during data reduction.
A spectrum of the internal wavelength calibration source
was obtained intermediate
among the galaxy spectra. HST's tracking was very stable during the observation. According
to the
jitter files reported by the Fine Guidance Sensor, the rms jitter during
the observation was 4.8 mas.

The initial steps of the data calibration (bias subtraction, dark rate subtraction, applying the
flatfield, and combining the two sub-exposures to reject cosmic-ray events)
were completed using the CALSTIS package in STSDAS.
The accuracy of the flatfield calibration was 1\%.
To reject most hot pixels from the data, we constructed a hot pixel mask
from dark frames 
obtained 15 hours before the M84 observations, and then interpolated across these
flagged pixels. This interpolation was performed along the dispersion axis, since the flux gradient
along the spatial axis near the nucleus is steep (because of the point source). The data were
wavelength calibrated and rectified by tracing the 
Ne and Cr emission lines (in the wavecal) for the dispersion axis and the nuclear point source for 
the spatial axis, and then applying these solutions for 
the geometric distortions to the data. The polynomial order of the fit along the dispersion axis
was kept low because the S/N in the nuclear point source at $\lambda \lesssim 3500$ \AA \
is very low (Fig.~1). At this point, spectra
from the first and second HST orbits were registered (by aligning on the continuum peak) and then
were combined, rejecting any remaining hot pixels. The count rate
in each pixel was calibrated to flux in absolute units by applying the standard sensitivity
curve for this grating obtained from the HST archive of calibration files. Fig.~1 shows
the central region of the reduced two-dimensional spectrum which we analyze in \S 3.

\subsection{Near-infrared Imaging}

On 21 July 1998, Quillen et al.~(1999) obtained near-infrared images
of M84 with the NICMOS Camera 2 (Thompson et al.~1998), 
which has a pixel scale of $0\farcs075$/pixel
(MacKenty et al.~1997).
Observations were obtained in each of three broadband filters F110W (1.10 $\mu$m),
F160W (1.60 $\mu$m), and F205W (2.05 $\mu$m),
corresponding roughly
to the J, H, and K bands respectively. The total integration time in each filter was 768 sec.
Quillen et al.~describe the observing procedure and data reduction only for the F160W image,
so we list the details they did not cover for the other bandpasses.
For the F205W image, the
accuracy of the flatfield is degraded toward the edge of the field of view, and 
we subtracted
the predicted background count rate (MacKenty et al.~1997), which is only a few percent of the 
count rate from the central few arcseconds of M84.
We adopted the flux calibration
and photometric zeropoint (i.e., the flux of Vega)
provided by the HST pipeline software. The internal photometric error is estimated to
be $\le$ 2\%. We did not attempt to transform the NICMOS
instrumental magnitudes to standard JHK magnitudes, because the color terms for
these observations of M84 in F110W and F205W 
are significant but are not calibrated very accurately. Since the purpose
of these near-IR images is to assess the internal extinction via a color-color plot, it
is sufficient to use the NICMOS instrumental system since the effective wavelengths
of our filters are known.
Fig.~2 shows the
near-IR images, which have a resolution (FWHM) of $0\farcs15$, and two near-IR color maps
[i.e., (J$-$H)$_{\rm N}$ and (H$-$K)$_{\rm N}$], which were formed from the ratios of F110W/F160W 
images and F160W/F205W images, respectively. The
subscript on these colors represents the fact that the magnitudes are in the NICMOS
system instead of the Johnson system. The (H$-$K)$_{\rm N}$ map shows false
color variations toward the edge of the field of view ($\approx 4''$ from the nucleus) symptomatic
of the lower quality of the flatfield in F205W. The (H$-$K)$_{\rm N}$ color values at pixels
within $0\farcs25$
of the nucleus have not been corrected for the effects of wavelength-dependent 
resolution. The resolution provided by NICMOS Camera 2 is
limited by diffraction at wavelengths greater than 1.75 $\mu$m (MacKenty et al.~1997), so the
nuclear point source is more extended in the F205W image than in the other bandpasses.

\section{Results}

As mentioned in \S 1, the nuclear point source is embedded in dust. 
We estimate this internal extinction in \S 3.1. In \S 3.2,
we correct for this extinction and analyze the nuclear spectrum.

\subsection{Estimating the Internal Extinction}

M84's nuclear region has been imaged in V and I (Paper~I) as well as J, H, and K
bands (this paper). We utilize the (V$-$H) map constructed by Quillen et al.~(1999), which
provides the longest wavelength baseline where the colors can be converted to the Johnson system. 
[As discussed in \S 2, the color term for converting fluxes in the F205W filter
to K band magnitudes is significant but poorly known. The color term for the F160W filter (H-band)
is insignificant relative to the internal error (which is the reason we drop the N subscript
from the (V$-$H) color.] However, estimating the extinction from one color
requires assuming the standard Galactic interstellar reddening curve. We test the validity of this 
assumption
by constructing a
color-color diagram from our near-IR images, which also provides a check on the extinction 
measurement indicated by the (V$-$H) map.  

From the (V$-$H), (J$-$H)$_{\rm N}$, and (H$-$K)$_{\rm N}$ color maps, we analyze 
only pixels lying
a distance $R$ from the nucleus such that $0\farcs25 \leq R \leq 1\farcs00$. This annulus excludes
pixels too close to the nucleus where the PSF is diffraction-limited for F205W, as
well as pixels affected by the dust lane passing $1\farcs5$ north of the nucleus. Thus, this 
annulus defines a region containing the nuclear dust 
and its surroundings that are not affected by the dust apparent in our images.
Fig.~3a show the histogram of (V$-$H) inside this aperture. The shaded region represents pixels
from the aperture lying along the major axis of
the nuclear dust (P.A. = $72\arcdeg$). Since the nuclear dust is roughly axisymmetric within this 
annulus, we make the simplifying
assumption that these points are representative of all dusty regions within the annulus. None of 
the shaded points 
lie toward the blue end of the histogram, suggesting that
this blue end represents the intrinsic color of the stellar population, and that the
rest of the histogram represents this color affected by internal reddening. The weighted average
of the shaded region is (V$-$H) = 3.50, and the weighted average of the two greatest bins
in the histogram (which we adopt as the intrinsinc color) is (V$-$H) = 3.08. The color excess is
then E(V$-$H) = 0.42, corresponding to $A_V = 0.50$. This assumes the standard Galactic reddening
curve (with $R \equiv A_V$/E(B$-$V) = 3.1) tabulated by Fitzpatrick (1999).

To determine if this assumed reddening curve is appropriate, we construct the (J$-$H)$_{\rm N}$ 
versus (H$-$K)$_{\rm N}$ color-color
plot for the same aperture (Fig.~3b). 
The colors in this plot define a band elongated parallel to the
reddening vector which we have determined for our three filters from the same Galactic reddening
curve adopted above. Thus, our adopted reddening curve
is justified. The apparent thickness of this band
is dominated by internal photometric error. The larger filled points mark the colors seen 
along the major axis of
the nuclear dust, i.e., the same region emphasized in Fig.~3a. These emphasized points in Fig.~3b
are concentrated toward the red end of the band, as are these points concentrated toward the
red end of the (V$-$H) histogram in Fig.~3a. This distribution in Fig.~3b is thus consistent
with our suggestion above that the bluer colors represent the intrinsic color of the stellar
population, and that the redder colors represent this intrinsinc color affected by internal 
reddening.
We measure the intrinsic colors of the stellar population to be (J$-$H)$_{\rm N}$ = 1.03 and 
(H$-$K)$_{\rm N}$ = 0.288, while the median of the dust major axis points has (J$-$H)$_{\rm N}$ = 
1.10
and (H$-$K)$_{\rm N}$ = 0.313. The color excesses are E(J$-$H)$_{\rm N}$ = 0.07 and
E(H$-$K)$_{\rm N}$ = 0.025, corresponding to $A_V = 0.48$. Within the errors this estimate agrees
with that from the (V$-$H) colors.

This measurement of $A_V$ could be underestimated. The intrinsic colors assumed in the analysis
above is actually not the intrinsic color of the stellar population if
there is a diffuse component of dust in M84. Such a dust component
is difficult to detect via broadband imaging, especially if its
distribution follows that of the starlight (e.g., Goudfrooij \& de Jong~1995). In Paper~I (see
their Fig.~4),
a histogram of (V$-$I) in the central
$14\farcs5 \times 8\farcs4$ showed a peak at (V$-$I) = 1.4. If this were to be adopted as the
intrinsic color to estimate $\left < A_V \right >$ (the mean extinction inside this
region), then the calculated dust mass (e.g.,
van Dokkum \& Franx 1995), from the pixels redder than the peak in that histogram, 
would be $\approx 1/7$ of the dust mass detected in the far-IR by IRAS (Roberts et al.~1991,
scaled to our adopted distance). Alternatively, Paper~I adopted (V$-$I) = 1.2 as the intrinsic
color, based on the mean for M84's
morphological type in Buta \& Williams~(1995), and the dust mass found agreed with the value based
on the far-IR flux. Therefore, the diffuse component of the dust should not be ignored.

Since we need the extinction toward the nucleus, it would be incorrect to calculate it
from the total dust mass
since this provides the extinction averaged over kpc scales. However, if  
E(V$-$I) = 0.2 (the difference between the peak (V$-$I) indicated by
Paper~I's WFPC2 images and the adopted intrinsic (V$-$I) based on morphological type) is
attributed to a diffuse component, then (based on the Galactic reddening curve) the extinction
associated with the putative diffuse dust component is $A_V = 0.38$. Therefore, for the
total extinction we adopt $A_V \approx 0.87$.

Does this estimated extinction, which we have derived from regions within the
$0\farcs25 \leq R \leq 1\farcs00$ annulus, apply to the nucleus as well? This depends
on the spatial distribution of the dust. The dust on this scale has the same morphology
as the ionized gas disk (Paper~I) which we know
to be associated with the nucleus (Paper~II). Thus, it is likely that the dust on this scale
also lies in a disk. Any variation in extinction across the dust would depend on the inclination
and thickness of the disk. For example, if the disk is viewed close to face-on or is thin
(compared with its diameter),
then the extinction toward the nucleus would be significantly less than it would be
toward larger radii, since the line of sight toward the nucleus would have less dust.
However, results from Paper~II suggest that neither of these scenarios is the case. The model fit 
to the dynamics of the gas disk indicates that the inclination is
$75\arcdeg - 85\arcdeg$ with respect to the plane of the sky. 
If the disk is indeed thin, one expects the
velocity dispersion in the gas to be small compared with the rotational velocity, yet Paper~II's
spectra exhibit very broad emission profiles even away from the nucleus (seen best in [S~II] 
$\lambda\lambda 6717, 6731$),
whose widths suggest $\sigma_{\rm gas} \sim 400$ km~s$^{-1}$. This is roughly equal to the
highest observed rotational velocity (even after deprojection). Therefore, we conclude that
the disk thickness is not negligible and that adopting this extinction estimate 
for the nucleus would not
be inappropriate.     

\subsection{The Nuclear Spectrum}

Since the extinction to the nucleus has been estimated, we turn our
attention to the STIS spectrum.
The long-slit spectrum in Fig.~1 shows strong absorption lines away from the nucleus,
including Fe I (5270~\AA),
the Mg I b lines at 5167~\AA, 5172~\AA, 5183~\AA \ (which appear blended together because
of the finite instrumental resolution and the stellar velocity dispersion), 
the G band attributed to CH (4300~\AA), and Ca II K and H (3933~\AA, 3968~\AA). The
emission lines seen (such as [O~II] $\lambda 3727$, H$\beta$, and [O~III] $\lambda\lambda
4959,5007$)
are strongly concentrated into a region $0\farcs28$ across centered on the nucleus.
To measure emission line fluxes, as well as the strengths of stellar absorption lines 
at the nucleus compared
with the strengths of those lines away from the nucleus, 
we must extract one-dimensional spectra on and off the nucleus.  

Fig.~4 shows the extracted spectra. The nuclear spectrum is an optimal extraction 
(i.e., each row is weighted
by its mean counts) of the central $0\farcs28$ along the slit.
To maximize the $S/N$ in the off-nuclear spectrum, we optimally extract two
spectra away from the nucleus centered at radius $R = +2\farcs6$ and
$-2\farcs6$ with size of $4\farcs8$, and then
add them together. The nuclear spectrum reveals several features not detected in the blue spectrum 
of
Ho, Filippenko, \& Sargent (1995) from
their ground-based spectroscopic survey of a sample of bright galaxies.
Given that the emission lines have relatively low equivalent widths (see below) and a
compact spatial distribution (Fig.~1), these emission
lines would be severely diluted by stellar light in ground-based spectra. The off-nuclear spectrum
shows absorption features typical of a stellar population whose light is dominated by
K giants, including the broad absorption feature of MgH centered at $\sim 5100$~\AA.
In contrast, the absorption features are weak in the nuclear spectrum. To determine if this
is simply a consequence of the expected higher velocity dispersion $\sigma$ near the nucleus, we
convolve the off-nuclear spectrum in Fig.~4 with a Gaussian broadening function whose width
is determined by the difference in quadrature
between the values of $\sigma$ at the nucleus and at the off-nuclear aperture.
Unfortunately, no measurements of the stellar kinematics in M84 with sub-arcsecond resolution
exist. We adopt $\sigma = 300$ km s$^{-1}$ for the off-nuclear aperture (Davies \& Birkinshaw 
1988). 
To estimate $\sigma$ at the nucleus at our resolution, we adopt the nuclear $\sigma = 600
\pm 37$ km s$^{-1}$ found
by
Kormendy et al.~(1996)
in the center of NGC~3115 after scaling to M84 to correct for the differences in galaxy distance
and presumed BH mass. 
This yields an estimate for M84 that $\sigma \approx 370$ km s$^{-1}$ at the nucleus.
The off-nuclear spectrum broadened to this estimated velocity dispersion is shown in Fig.~4 
as a dotted line. The absorption
lines in this spectrum are much stronger than those in the nuclear spectrum, implying that
the weakness of the stellar absorption lines at the nucleus is not an artifact of the
stellar kinematics. Very
weak features are present in the nuclear spectrum at the wavelengths of Mg I b and the G band, but 
these features are real with only at most $2\sigma$
confidence. The nuclear spectrum has a clear absorption feature from MgH near 5100~\AA, while the
Ca II K and H (3933~\AA, 3968~\AA) absorption lines 
are probably not stellar in origin (see below).
This exercise suggests that the nuclear continuum (which Fig.~4 shows to be moderately bluer than 
that in the
off-nuclear spectrum) is dominated by
a featureless spectrum from an AGN or hot stars. The contribution from cool stars is
$\sim 25$\% at 5200~\AA \ (from the diminished equivalent width of Mg I b), and is 
insignificant in the blue where this other continuum component dominates.
In \S 4 we will address the nature of the bluer continuum component using the emission lines and 
the spectral energy distribution.
 
The Ca II K and H (3933~\AA, 3968~\AA) absorption lines (Fig.~4) 
are clearly much narrower in the nuclear spectrum than in the broadened off-nuclear spectrum. 
The profile widths in the nuclear spectrum are unresolved,
indicating that the absorption arises in a dynamically cold region where the velocity dispersion
of the absorbing gas is small relative to its rotational velocity.  
These lines therefore are produced by interstellar absorption in the gas associated with 
the dust that
extends in front of the nuclear point source. The dust 
within a projected distance of $\approx 80$
pc of the nucleus is likely to be associated with the nucleus, 
rather than lying in the
foreground (see \S 3.1). Although the extracted spectrum covers $0\farcs28$
(23 pc) along the slit, we would not expect to see rotational broadening
in these line profiles because the absorbing gas is illuminated by a point source.
Confirmation
that the dust does indeed lie in a disk requires detecting the molecular gas associated with the 
dust and measuring
its kinematics.

We can use the equivalent widths of these Ca II K and H absorption lines in the nuclear spectrum
(9.0~\AA \ and 4.5~\AA, respectively) for a very rough consistency check on the extinction 
derived in \S 3.1. 
These line profiles show no evidence of saturation, so
we estimate the column density of Ca II in the gas phase along the line of sight toward the
nucleus (following Spitzer 1978) to be $N$(Ca II) = $9.5 \times 10^{13}$ cm$^{-2}$.
Unfortunately, the abundance of Ca II in the interstellar medium is not correlated 
with the gas density of atomic and molecular hydrogen (which is related to the extinction through
the gas to dust ratio) because its abundance can be altered significantly by 
the two competing effects of
condensation of Ca II
onto dust grains and the destruction of dust grains
(e.g., Savage \& Mathis 1979). Thus, we cannot directly convert the Ca II column density into
a value of $A_V$. Instead, as a surrogate we can utilize Na I, since its abundance is correlated 
with that
of H I + H$_2$ (Welsh et al.~1997), and can find the ratio of Na I to Ca II column densities
required by our assumed value of $A_V$ and $N$(Ca II). 
However, even this is problematic because the STIS red spectrum from Paper II did not cover 
the Na I D lines. Nevertheless, $A_V \approx 0.87$ and our observed value of $N$(Ca II)
imply that $N$(Na I)/$N$(Ca II)
= 0.1 assuming the Galactic gas to dust ratio (Savage \& Mathis 1979).
For the 64 Galactic lines of sight studied by Siluk \& Silk (1974), the value of
$N$(Na I)/$N$(Ca II) ranges from $< 0.05$ to 90. If the range of abundances in M84's
interstellar medium is similar to those
in the Galaxy,
then $A_V \approx 0.87$ toward the nucleus is not
inconsistent with the equivalent widths of the Ca II absorption lines.

Since our main purpose is to determine the ionization mechanism, we need to measure the emission 
line fluxes for [O~II] $\lambda 3727$,
[O~III] $\lambda 5007$, and H$\beta$. These line ratios will then be considered with
the fluxes for [O~I] $\lambda 6300$, [N~II] $\lambda 6583$, and 
[S~II] $\lambda 6717 +
\lambda 6731$ relative to H$\alpha$ from the nuclear spectrum in Paper~II (see their
Fig.~3 for $R = 0\farcs00$). Since the blue emission lines are seen only within the central
$\approx 0\farcs28$,
we are able to place useful constraints on the ionization mechanism only at the
nucleus, not at larger radii in the 82 pc diameter gas disk. These emission line ratios clearly 
distinguish normal H II regions,
LINERs, and Seyfert galaxies 
(e.g., Veilleux \& Osterbrock 1987). The emission line ratios of
[O~III] $\lambda 5007$/H$\beta$, [O~I] $\lambda 6300$/H$\alpha$, [N~II] $\lambda 6583$/H$\alpha$,
and ([S~II] $\lambda 6717 +
\lambda 6731$)/H$\alpha$ are insensitive to uncertainties in the reddening correction, unlike other 
ratios such as [O~II] $\lambda 3727$/[O~III] $\lambda 5007$. This advantage is particularly 
important
for M84 since our knowledge of the reddening is uncertain. However, a significant disadvantage 
(which we address in \S 4) in the use
of Balmer emission line fluxes is that their apparent fluxes can be significantly lower than their 
true
fluxes
if the stellar light in the aperture has strong Balmer absorption lines, which could arise if
there is a significant population of A stars at the nucleus.

Fig.~4b shows the nuclear spectrum corrected for the internal extinction of $A_V = 0.87$
estimated above. The emission line profiles have a velocity
structure more complicated
than a Gaussian profile, reflecting the velocity gradient along the slit seen in the 
emission lines in Fig.~1. Consequently, instead of fitting Gaussian profiles to the emission lines,
the continuum flux, line flux, equivalent width, and width of each line,
shown in Table~1, were measured directly from the spectrum. We quote the fluxes corrected for 
Galactic
extinction only, as well as fluxes corrected for Galactic and internal extinction.   
The [O~III] profile was deblended by
assuming that the ratio of their fluxes F([O~III] $\lambda 4959$)/F([O~III] $\lambda 5007$) 
is 0.335, as given by the ratio of
their transition probabilities. The line widths have been corrected for the instrumental width
expected for a point source (\S 2) by assuming that the observed width is the sum in
quadrature of the true and instrumental widths.

Measuring emission line fluxes for [O~I] $\lambda 6300$, H$\alpha$, [N~II] $\lambda 6583$, 
and [S~II] $\lambda
\lambda 6717, 6731$ from Paper~II's spectrum is more uncertain. Even though this
spectrum has higher resolution, the line profiles (especially [N~II] and H$\alpha$)
are severely blended. Since we could find no clear method to deblend the profiles uniquely,
we assume that the peak flux above the continuum level at the central wavelength of each emission 
line is
representative of that emission line's total flux. Since it is difficult to locate the
peak of the [N~II] $\lambda 6548$ profile, we assume that 
F([N~II] $\lambda 6548$)/F([N~II] $\lambda 6583$) = $1/3$, which is the ratio of their
transition probabilities.
Our assumption that the peak flux is representative of the total flux implies that the
velocity structure of each emission line is identical, which might not be strictly true.
M84's nucleus might contain a broad line region, in which case the H$\alpha$ profile would not be
identical to those of the forbidden lines. However, 
any broad H$\alpha$ component must be at most a small fraction of the total H$\alpha$ flux, because
a broad H$\beta$ component is not obvious. Also, the forbidden lines might have different
velocity widths, since their critical densities for collisional de-excitation are different.
Correlations between line width and critical density have been found in Seyfert galaxies
(e.g., De Robertis \& Osterbrock~1984, 1986) and LINERs (e.g., Filippenko \& Halpern~1984).
Table~2
lists our measured line fluxes relative to H$\alpha$, separately corrected for Galactic reddening
only, as well as Galactic and internal reddening.
The errors in these fluxes reflect the fact that the observed line profiles cannot be decomposed
uniquely, and are dominated by the uncertainty in our assumption that the velocity
structure in each component is identical. We conservatively adopt errors of 50\% for 
[N~II] $\lambda 6583$/H$\alpha$ and 25\% for [O~I] $\lambda 6300$/H$\alpha$ and 
([S~II] $\lambda 6717 + \lambda 6731$)/H$\alpha$. 
We have much more
confidence in the [O~I] and [S~II] relative fluxes compared with
the [N~II] relative flux because these
lines are not blended with any other emission lines, so only the uncertainty in H$\alpha$ is
present for these two line ratios.

\section{Interpretation}

The goal of this section is to determine the constraints that can be placed on the nuclear
emission mechanism from the nuclear emission line ratios and spectral energy distribution.
The emission line ratios (discussed in \S 4.2) include the fluxes in Balmer emission lines 
measured in \S 3.2.
These fluxes might be underestimated if there is significant Balmer absorption arising
from the stellar population at the nucleus. If
significant, such absorption lines could cause an underestimate of the H$\alpha$ and H$\beta$ 
emission
line flux, leading to a misclassification of the emission line spectrum.
We consider this
possibility in \S 4.1.

\subsection{The Insignificance of Balmer Absorption Features}

In order to use the line fluxes as a diagnostic of the ionization mechanism at the
nucleus, we must 
determine
the significance of Balmer absorption lines in the nuclear spectrum. 
Although the nuclear spectrum in
Fig.~4b shows that H$\beta$ and H$\gamma$ are in 
emission, the nebular emission component for higher Balmer transitions is not observed.
We would expect the higher transitions to
appear in absorption, if A stars are significant. However, the [S~II] $\lambda\lambda
4069, 4076$
emission dilutes any H$\delta$ $\lambda 4101$ absorption, and the Ca II H $\lambda 3968$ absorption 
from 
the dust disk masks any H$\epsilon$ $\lambda 3970$ absorption. Even higher Balmer transitions
have lower equivalent widths in A stars, rendering their detection more difficult. The best 
spectral
feature in A stars remaining is the prominent Balmer absorption edge at 3646~\AA.
Although the nuclear spectrum in Fig.~4b shows no evidence of a Balmer absorption edge,
the $S/N$ in the continuum at the wavelength of this absorption edge is only 10 pixel$^{-1}$.
It could be that a Balmer absorption edge is indeed present, but is significantly diluted by a 
featureless continuum
that may be emitted by the nucleus. 
We construct synthetic spectra and compare them with the data to place an upper limit on
the possible contribution of Balmer absorption lines (primarily from A stars). 

Our synthetic spectra are simply linear combinations of an A0~V spectrum from the library
of Pickles~(1998) with a power law spectrum. The A0~V spectrum has a resolution of
5~\AA \ (which is very close to our resolution) and has solar metallicity. 
The power law component adopted has $F_{\lambda} \propto \lambda^{\beta}$ (where
$\beta = 4.33$ is a good approximation of the shape of the continuum in the nuclear spectrum
at $\lambda \leq 5000$~\AA \ (Fig.~4b). We adopt an A0~V spectrum as a proxy for a hypothetical 
nuclear cluster of
young stars
in favor of a more complicated approach
involving synthetic starburst spectra (e.g., Bruzual \& Charlot~1993;
Leitherer \& Heckman~1995),
because starburst models involve several parameters
which would be very poorly constrained in this case. This includes the star formation
history (i.e., an instantaneous starburst or a constant star formation rate), age, initial mass
function, and metallicity. In utilizing more sophisticated starburst models, it would be very 
difficult to translate 
the observed lack of a Balmer
absorption edge in M84's nucleus to a meaningful limit on possible Balmer absorption line
equivalent width.

Fig.~5 shows the input 
A0~V 
and power law spectra, where both have been normalized to unit flux at 5556~\AA,      
and the synthetic spectra constructed from a linear combination of these two input spectra
with Gaussian noise added to simulate the observed nuclear continuum component from Fig.~4b. 
The fit to the
continuum of the observed spectrum is superimposed for comparison. If A stars represent
at least 15\% of the light in the nuclear point source, then we should have detected
a Balmer absorption edge. We conclude
that A0 V stars contribute at most 10\% of the total light in the blue nuclear spectrum, and
the equivalent width (EW) of Balmer absorption features are at most 10\% of the
strengths in an A0 V star,
for which Pickles~(1998) finds H$\beta$ and
H$\alpha$ equivalent widths to be 13.4~\AA \ and 9.8~\AA, respectively. This implies
that the upper limits to absorption lines are EW(H$\beta$) $< 1.4$~\AA \ and
EW(H$\alpha$) $< 1.0$~\AA.

Given the continuum flux near H$\beta$ corrected for internal extinction listed in Table~1, this 
upper limit on H$\beta$ absorption
could hide an emission line flux of $< 3.8 \times 10^{-16}$ erg cm$^{-2}$ s$^{-1}$, so the
H$\beta$ flux, corrected for possible underlying stellar absorption,
is at most $2.49 \times 10^{-15}$ erg cm$^{-2}$ s$^{-1}$, or at most 18\% above the observed flux.
For the correction of H$\alpha$, the equivalent width of the H$\alpha$ + [N~II] profile
in Paper~II's spectrum is 274~\AA. Table~2 indicates that the H$\alpha$ flux is 38\% of the
H$\alpha$ + [N~II] flux, so EW(H$\alpha$) is approximately 105~\AA. Any correction for
possible stellar absorption (which is at most 1\%) is much smaller than the uncertainty in the
H$\alpha$ flux arising through the uncertainty in emission line deblending. Consequently,
we ignore this correction.

\subsection{Emission Line Diagnostics of the Ionizing Continuum}

Fig.~6 shows the emission line ratios in the diagnostic diagrams from Veilleux \& Osterbrock 
(1987),
which are insensitive to uncertainties in the reddening correction, as well as the [O~II] 
$\lambda 3727$/[O~III] $\lambda 5007$ diagram (e.g., Shields \& Filippenko~1990).
The point representing M84's nucleus is clearly segregated from the regime of normal H II regions. 
It
is also not consistent with the H II region models of Filippenko \& Terlevich~(1992) in which O 
stars photoionize solar metallicity nebular gas
and the ionization parameter is low
($\log U \approx -3.7$ to $-3$). Their Fig.~1 shows that these models produce LINER spectra
with lower excitation 
than we find in M84's nucleus, as indicated by the value of [O~III] $\lambda 5007$/H$\beta$.   
Our measurements indicate that hot stars play at most a very minor role in the ionization mechanism
at the nucleus. 
This would be consistent with our determination that A stars do not significantly
contribute to the spectrum of the nucleus, if
the IMF of the stellar population associated with the nucleus is roughly
similar to a Salpeter function.
In the emission line ratio diagrams, for convenience we have adopted lines that separate the
loci of Seyfert galaxies and LINERs given by [O~III] $\lambda 5007$/H$\beta$ = 3 and 
[O~II] $\lambda 3727$/[O~III] $\lambda 5007$ = 1.
M84's nuclear line ratios are very close to these lines in all four plots,
so its classification between these two types is ambiguous. However, exact divisions between
these two types are somewhat arbitrary.
We compare the position of M84 in these
plots to that of M87 [whose nuclear and off-nuclear line ratios are taken from Chakrabarti~(1995) 
and 
Dopita et al.~(1997)], since these two galaxies have many features in common. Both are radio
galaxies containing nuclear gas disks (with diameters $\sim 100$ pc). In each case, the
gas disk has a Keplerian velocity field (Paper~II; Macchetto et al.~1997), indicating the presence 
of a $\sim 10^9 \ M_{\odot}$ dark compact object, which presumably is in the form of a black hole. 
M84's nucleus clearly has significantly higher
excitation than most LINERs, including M87. It is more consistent with the higher excitation found
in Seyfert nuclei. The [O~II] $\lambda 3727$/[O~III] $\lambda 5007$ ratio, which is sensitive
to the assumed reddening correction, is marginally 
more consistent with the classification derived from the
other line ratios shown (which are insensitive to reddening) if there is no correction
for internal reddening at all. Perhaps the amount of dust within a few pc of the nucleus is
lower than it is at larger radii in the dust disk, contrary to our expectation given in \S 3.1.
However, the extinction to the nucleus cannot be zero, since Ca II absorption lines
from the ISM are seen against the background nuclear point source.

\subsection{The Ionizing Photon Flux from the Nuclear Source}
 
Since the ionization conditions in M84's nucleus are more consistent with Seyfert nuclei
than LINERs, the ionization mechanism is probably photoionization from either the AGN
or perhaps high-velocity shocks [which can generate a strong
UV radiation field capable of ionizing the gas, resulting in a high-excitation emission line
spectrum (e.g., Allen et al.~1998)].
The ionizing flux required can be estimated by assuming case B
recombination (i.e., the
gas is optically thick to Lyman continuum photons) and that
the electron temperature in the ionized gas is $10^4$ K. The ionizing luminosity $Q({\rm H}^0)$ 
is then related to the H$\beta$ luminosity 
(Osterbrock 1989).
The H$\beta$ flux from Table~1 
corresponds to a luminosity $L({\rm H}\beta)$ of $2.9 \times 10^{37}$ erg s$^{-1}$ 
(corrected for Galactic extinction but not internal extinction) or $7.4 \times
10^{37}$ erg s$^{-1}$ (corrected for both Galactic and internal extinction).
We calculate that the ionizing luminosity is $Q({\rm H}^0) \approx
1.5 \times 10^{50}$ photons s$^{-1}$. This is a lower limit
if the nebular gas close to the nucleus (within the volume spatially unresolved in the
spectrum) has a clumpy distribution so that some ionizing photons can escape without
being reprocessed by the gas. 

\subsection{The Spectral Energy Distribution of the Nuclear Source}

The spectral energy distribution (SED) of the nonstellar point source in the nucleus of M84
is important because knowledge of the SEDs in low luminosity AGN should
provide key insights into the physics of these objects. However, the measurement of fluxes from low
luminosity AGN is complicated by the need for both high sensitivity and spatial resolution, to
distinguish the faint AGN from its host galaxy.
Ho (1999) compiled such data from the literature for a sample of seven low luminosity AGN including
M84. This preliminary SED for M84 can be improved significantly through the addition of 
measurements
from our new HST data and from archival Infrared Satellite Observatory (ISO) images.

Our STIS spectrum provides a clear detection of the nuclear point source extending blueward
of the V band, a region for which no detections existed previously in its SED (Ho 1999). 
To determine if the nuclear source is variable, in Table~3 we list the nuclear
V band magnitudes (corrected for Galactic extinction) determined from our STIS spectrum 
(Fig.~4a) and the
two epochs of HST V band images. 
For the two HST V band images, we isolate the nuclear
flux from the host galaxy by fitting model PSFs (Krist \& Hook 1997) to the nucleus (as in Paper 
I).
From our STIS spectrum, we extract the intensity profile along the slit (perpendicular to the
dispersion axis) using a bandpass whose wavelength coverage matches the
filter bandpasses of the HST V band images. We then repeat the PSF fitting (albeit with one 
spatial dimension
instead of two) to isolate the nuclear
flux from the host galaxy, finding that the observed flux at
the V band is $5.1 \times 10^{-17}$ erg cm$^{-2}$ s$^{-1}$ \AA$^{-1}$. 
The error on these magnitudes of the nuclear source is no larger than 0.1 mag, which is
dominated by the uncertainty in isolating the nucleus from the host galaxy. We list the nuclear 
magnitudes from these
three epochs in Table~3. During this time, the nucleus has brightened in V band by $\approx
75$\%, adding more confidence to our result that the nucleus is nonstellar.
  
To measure the flux from the nuclear point source in the NICMOS images in Fig.~2, first
we subtract the host galaxy using
a fit to the surface brightness distribution
beyond $0\farcs6$ from the nucleus. This restriction is required to avoid 
emission from the nuclear point source within the first diffraction ring. 
The fit was extrapolated inward from this 
distance toward the nucleus. Since the stellar surface brightness profile is very shallow within 
its
break radius of $2\farcs45$
(Quillen et al.~1999), this extrapolation is insensitive to the details of the fit. 
Subtracting this model for the host galaxy from the original
image isolates the nuclear point source. We then measured the flux of
this source using aperture corrections derived from a model PSF generated with Tiny Tim
(Krist \& Hook 1997). Our measured fluxes are 0.22 mJy (F110W), 0.41 mJy (F160W), and
0.42 mJy (F205W).

The SED can be extended into the mid-infrared using observations
by ISO\footnote{The Infrared Space Observatory (ISO) is an ESA
(European Space Agency) mission with the participation of ISAS (Japan) and NASA (USA).}
(Kessler et al.~1996) with ISOCAM (Cesarsky et al.~1996)
obtained by F. Macchetto on 5 July 1996. The images were obtained in
the three bandpasses LW2 (6.75 \micron), LW7 (9.62 \micron), and
LW3 (15 \micron) in the 3$''$/pixel mode (spatial resolution $\approx 5'' - 8''$). 
We retrieved these images
from the ISO archive and then resampled the WFPC2 F547M image,
after subtracting the nuclear point source,
to match the plate scale and resolution of each of the three ISOCAM images.  
No significant difference was found between the light profiles in 
this resampled WFPC2 image (which provides
the stellar light distribution)
and the ISOCAM images, implying that the mid-infrared
emission is associated with the stars and that the nuclear point source was not
detected at these mid-infrared wavelengths. For each
ISOCAM image, we estimated an upper limit on the flux from this nuclear source (detected
in the WFPC2 and NICMOS images) from twice the flux of the brightest ISOCAM pixel (minus the
background). These upper limits are $< 10$ mJy (LW2), $< 6$ mJy (LW7), and $< 5$ mJy (LW3).

Table~4 compiles these new flux measurements with the previous measurements from Ho (1999) that
have not been superseded by new results. We note that Ho (1999) misquoted the 2300~\AA \ upper 
limit
found by Zirbel \& Baum (1998). This updated SED is shown in Fig.~7 along with our spectrum
of the nucleus from Fig.~4b. We note the caveat that the observations composing this SED are 
obviously not
simultaneous, and the source is variable at least at optical wavelengths (see above). 
As noted in \S 3.2, the contribution of host galaxy starlight to the nuclear spectrum is
$\sim$ 25\% of the total light at 5200~\AA \ but is negligible at shorter wavelengths. 
We superimpose the mean SED of
radio loud quasars (Elvis et al.~1994) and the SED of BL Lac (Bregman et al.~1990),
both normalized to M84 at K band. More recent compilations of the SEDs of BL Lac objects
in the literature
do not have sufficient frequency resolution to compare them adequately with M84. 
We do not overplot the mean SED of Seyfert 2 nuclei
(e.g., Schmitt et al.~1997) because it is approximately two orders of magnitude too faint
at radio frequencies. M84's SED is significantly different from the mean radio loud quasar
SED at the wavelength range covered by our spectrum. At blue and near-UV wavelengths,
M84's SED is very red, whereas the mean radio loud quasar SED has the big blue bump. M84's SED
is more similar to that of BL Lac. This normalization of BL Lac's SED has an ionizing luminosity
of $Q({\rm H}^0) = 1.4 \times 10^{50}$ erg s$^{-1}$. This is only 7\% lower than the lower limit
on $Q({\rm H}^0)$
for M84's nucleus estimated from
the H$\beta$ luminosity in \S 4.3, suggesting that the SEDs of M84's nucleus and BL Lac could very
well be similar at energies greater than 13.6 eV ($3.28 \times 10^{15}$ Hz).  Since our M84 SED has 
sparse frequency coverage, 
we use the normalized BL Lac SED shown in
Fig.~7 as a template to estimate that the bolometric luminosity of M84's nuclear point source
is $\sim 1 \times 10^{41}$ erg s$^{-1}$, assuming isotropic emission. However, as a cautionary 
note,
this could be significantly
underestimated if the nuclear emission is beamed away from us (see \S 5). Since the mass of the 
compact dark object (most likely
a BH) is $\approx 1.5 \times 10^9 \ M_{\odot}$ (Paper II), this estimated bolometric luminosity is 
only
$\sim 5 \times 10^{-7}$ of the Eddington luminosity, implying that the BH in M84 is accreting at a 
very low
rate.

\section{Conclusion}

The following four implications of our observations show that the nuclear point source in M84 
cannot be
a star cluster:

\par\noindent
1. Stellar absorption lines at the nucleus are very weak. Cool stars make a
minor contribution to the total light.

\par\noindent
2. A Balmer absorption edge is not detected. Thus, the contribution of A stars is insignificant.

\par\noindent
3. The emission line ratios indicate that the gas is not photoionized by hot stars. Therefore,
OB stars are insignificant as well.

\par\noindent
4. The spectral energy distribution (SED) of the nuclear source is also strong evidence that
it is nonstellar. We have updated the preliminary SED compiled by Ho (1999).
This updated SED is roughly flat, which is vastly different from the SED 
of a stellar cluster. Among types of AGN, the SED of M84's nucleus is more similar to the SED of BL 
Lac than that of radio loud quasars. Specifically,
M84's nucleus is characterized by a very red continuum at optical to UV wavelengths
where radio loud quasars have very blue continua. We have adopted the SED of BL Lac as a template
to estimate the bolometric luminosity of M84's nuclear source assuming isotropic emission, 
implying that the luminosity
is only $\sim 5 \times 10^{-7}$ of the Eddington luminosity.

The similarity between the SEDs of M84's nuclear source and BL Lac plus its V band
variability suggest the possibility that M84 might be
a misaligned BL Lac object. In that case, our assumption of isotropic emission, for the purpose
of estimating the bolometric luminosity, would be invalid since the radiation could be beamed
away from us. The characteristics of M84's radio source imply that this possibility should
be investigated further. Its radio jets
are nearly in the plane of the sky (as indicated by the roughly comparable power in the radio jets
and lobes on either side of the nucleus; Laing \& Bridle 1987). Is the optical emission beamed
in the same manner? If so, then the evidence for
M84 being a misaligned BL Lac object would be strengthened. 
This would fit with the evidence
suggesting that BL Lac objects are a subset of FR I radio galaxies that are oriented such
that their jets are beamed
directly at us (e.g., Urry \& Padovani 1995). In that case, it is interesting that Centaurus~A and
M87 (two other nearby FR I radio
galaxies) also show evidence of being misaligned BL Lac objects. 
For Centaurus~A where direct observation of the nucleus is not possible at optical and UV
wavelengths,
Morganti et al.~(1991, 1992) found that the ionization conditions in gaseous filaments $\sim
9$ kpc northeast of the nucleus suggest that this gas is photoionized by beamed radiation from the 
nucleus.
However, the shock models constructed by Sutherland, Bicknell, \& Dopita (1993) suggest that
the ionization source of these clouds could arise from the interaction between the radio jet and
the galaxy's ISM. For M87, the evidence of a misaligned BL Lac is indicated by its variability
characteristics at optical wavelengths (Tsvetanov et al.~1998), 
the detection of superluminal motion in its relativistically
boosted jet (Biretta, Zhou, \& Owen 1995; Biretta, Sparks, \& Macchetto 1999), 
and its radio to optical and optical to 
X-ray spectral indices (Biretta, Stern, \& Harris 1991; Boksenberg et al.~1992).

\acknowledgments

We thank the referee for several useful suggestions.
This work was supported by NASA Guaranteed Time Observer funding to the STIS 
Science Team and by funding from NASA through grant number
GO-07868.01-96A (to ACQ)
from the Space Telescope Science Institute, which is operated by
AURA, Inc., under 
NASA
contract NAS5-26555.


\clearpage

\par\noindent
{\bf References}

\par\noindent
Allen, M.~G., Dopita, M.~A., \& Tsvetanov, Z.~I.~1998, \apj, 493, 571\\
Biretta, J.~A., Sparks, W.~B., \& Macchetto, F.~1999, \apj, 520, 621\\
Biretta, J.~A., Stern, C.~P., \& Harris, D.~E.~1991, \aj, 101, 1632\\
Biretta, J.~A., Zhou, F., \& Owen, F.~N.~1995, \apj, 447, 582\\
Boksenberg, A., et al.~1992, \aap, 261, 393\\
Bower, G.~A., Heckman, T.~M., Wilson, A.~S., \& Richstone, D.~O.~1997, \apj,
483, L33\\ 
\indent (Paper~I)\\
Bower, G.~A., et al.~1998, \apj, 492, L111 (Paper~II)\\
Bregman, J.~N., et al.~1990, \apj, 352, 574\\
Bruzual, G.~A., \& Charlot, S.~1993, \apj, 405, 538\\
Burstein, D., \& Heiles, C. 1984,
\apjs, 54, 33\\
Buta, R., \& Williams, K.~L.~1995, \aj, 109, 543\\
Cesarsky, C.~J., et al.~1996, \aap, 315, L32\\
Chakrabarti, S.~K.~1995, \apj, 441, 576\\
Davies, R.~L., \& Birkinshaw, M.~1988, \apjs, 68, 409\\
De Robertis, M.~M., \& Osterbrock, D.~E.~1984, \apj, 286, 171\\
De Robertis, M.~M., \& Osterbrock, D.~E.~1986, \apj, 301, 727\\
Dopita, M.~A., et al.~1997, \apj, 490, 202\\
Elvis, M., et al.~1994, \apjs, 95, 1\\
Fabbiano, G., Kim, D.-W., \& Trinchieri, G.~1992, \apjs, 80, 531\\
Faber, S.~M., et al.~1997, \aj, 114, 1771\\
Filippenko, A.~V., \& Halpern, J.~P.~1984, \apj, 285, 458\\
Filippenko, A.~V., \& Terlevich, R.~1992, \apj, 397, L79\\
Fitzpatrick, E.~L.~1999, \pasp, 111, 63\\
Goudfrooij, P., \& de Jong, T.~1995, \aap, 298, 784\\
Harms, R.~J., et al.~1994, \apj, 435, L35\\
Ho, L.~C.~1999, \apj, 516, 672\\
Ho, L.~C., Filippenko, A.~V., \& Sargent, W.~L.~W.~1995, \apjs, 98, 477\\
Jaffe, W., Ford, H.~C., O'Connell, R.~W., van den Bosch, F.~C., \&
Ferrarese, L.~1994, \aj,\\ 
\indent 108, 1567\\
Jones, D.~L., Terzian, Y., \& Sramek, R.~A.~1981, \apj, 246, 28\\
Kessler, M.~F., et al.~1996, \aap, 315, L27\\
Kimble, R.~A., et al.~1998, \apj, 492, L83\\
Kormendy, J., et al.~1996, \apj, 459, L57\\
Krist, J., \& Hook, R.~1997, The Tiny Tim User's Guide, 
http://scivax.stsci.edu/$\sim$krist/tinytim.html\\
Laing, R.~A., \& Bridle, A.~H.~1987, \mnras, 228, 557\\
Leitherer, C., \& Heckman, T.~M.~1995, \apjs, 96, 9\\
Macchetto, F., et al.~1997, \apj, 489, 579 \\
MacKenty, J.~W., et al.~1997, NICMOS Instrument Handbook, Version 2.0
(Baltimore:\\ 
\indent STScI)\\
Maoz, D., et al.~1998, \aj, 116, 55\\
Morganti, R., et al.~1991, \mnras, 249, 91\\
Morganti, R., Fosbury, R.~A.~E., Hook, R.~N., Robinson, A., \& Tsvetanov, Z.~1992,
\mnras,\\
\indent 256, P1\\
Mould, J., et al.~1995, \apj, 449, 413\\
Osterbrock, D.~E.~1989, Astrophysics of Gaseous Nebulae and Active 
Galactic Nuclei (Mill\\ 
\indent Valley, CA: University Science Books)\\
Pickles, A.~J.~1998, \pasp, 110, 863\\
Quillen, A.~C., Bower, G.~A., \& Stritzinger, M.~1999, \apjs, in press (astro-ph/9907021)\\
Richstone, D., et al.~1998, \nat, 395, A14\\
Roberts, M.~S., Hogg, D.~E., Bregman, J.~N., Forman, W.~R., \& Jones, C.~1991,
\apjs, 75,\\ 
\indent 751\\
Savage, B.~D., \& Mathis, J.~S.~1979, \araa\, 17, 73\\
Schilizzi, R.~T.~1976, \aj, 81, 946\\
Schmitt, H.~R., Kinney, A.~L., Calzetti, D., \& Storchi-Bergmann, T.~1997, \aj, 114, 592\\
Shields, J.~C., \& Filippenko, A.~V.~1990, \aj, 100, 1034\\
Siluk, R.~S., \& Silk, J.~1974, \apj, 192, 51\\
Spitzer, L.~1978, Physical Processes in the Interstellar Medium (New York: Wiley)\\
Sutherland, R.~S., Bicknell, G.~V., \& Dopita, M.~A.~1993, \apj, 414, 510\\
Thompson, R.~I., Rieke, M., Schneider, G., Hines, D.~C., \& Corbin, M.~R.~1998,
\apj, 492,\\ 
\indent L95\\
Tsvetanov, Z.~I., et al.~1998, \apj, 493, L83\\
Urry, C.~M., \& Padovani, P.~1995, \pasp, 107, 803\\
van Dokkum, P.~G., \& Franx, M.~1995, \aj, 110, 2027\\
Veilleux, S., \& Osterbrock, D.~E.~1987, \apjs, 63, 295\\
Walborn, N., \& Baum, S.~(ed) 1998, STIS Instrument Handbook, Version 2.0 
(Baltimore:\\ 
\indent STScI)\\
Welsh, B.~Y., Sasseen, T., Craig, N., Jelinsky, S., \& Albert, C.~E.~1997, \apjs,
112, 507\\
Woodgate, B.~E., et al.~1998, \pasp, 110, 1183\\
Zirbel, E.~L., \& Baum, S.~A.~1998, \apjs, 114, 177

\clearpage

\figcaption{The reduced two-dimensional spectrum of M84. The wavelength scale is
in the observer's reference frame. The radius scale is defined such that $R=0$
corresponds to the position of the nucleus, and positive radii correspond to the
direction along the slit with P.A. $= 130\arcdeg$ on the sky. The dynamic range
of the linear intensity scale covers $-1.0$ (white) to $\geq 3.0$ (black) in units of 
$10^{-17}$ erg cm$^{-2}$ s$^{-1}$ \AA$^{-1}$. 
\label{fig1}}

\figcaption{The WFPC2 image (taken from Paper~I) and NICMOS images of the nuclear 
region of M84.
All images have identical field of view and orientation (north is up, and east to the
left), with the nucleus at the origin of the coordinate system. 
The top four panels show the images in the indicated bandpasses (the
intensity stretch is logarithmic). The displayed intensity ranges and peak intensities
(in units of erg cm$^{-2}$ s$^{-1}$ \AA$^{-1}$ arcsec$^{-2}$)
for the WFPC2/F547M image are: $2.4 \times 10^{-16}$ to $2.4 \times 10^{-15}$, and 
$3.1 \times 10^{-15}$, respectively.
The displayed intensity range for all three NICMOS images is 1.24 mJy arcsec$^{-2}$ to 12.4 
mJy arcsec$^{-2}$, and their peak intensities in these units are 9.4 (F110W), 13.7 (F160W),
and 12.3 (F205W).
The lower
two panels show
(J$-$H)$_{\rm N}$ and (H$-$K)$_{\rm N}$.
The displayed color values 
range linearly from (J$-$H)$_{\rm N}$ of 1.18 to 0.99 and (H$-$K)$_{\rm N}$ of
0.42 to 0.19. Darker shades correspond to redder colors. Since the PSF through the F205W
filter is different from that through F110W and F160W, the apparent (H$-$K)$_{\rm N}$
within $0\farcs25$ of the nuclear point source should be disregarded (see \S 3.1).
\label{fig2}}

\figcaption{Analysis of the colors (corrected
for foreground Galactic reddening) inside an annular aperture centered on the nucleus (see \S 3.1).
(a) The distribution of (V$-$H). The shaded region represents pixels lying in the major axis
of the nuclear dust lane. (b) 
The (J$-$H)$_{\rm N}$ versus (H$-$K)$_{\rm N}$ color-color plot. The larger filled circles 
represent
pixels lying in the major axis of the nuclear dust lane. The open boxes represent
the median of these twenty points and the adopted color unreddened by dust apparent
in the broadband images. The open circle shows the color in an aperture with
radius of $0\farcs25$ centered on the nucleus. \label{fig3}}

\figcaption{(a) Extracted spectra (solid lines) of the nuclear point source and the off-nucleus
stellar light. The flux scale on the left is in units of $10^{-15}$ erg cm$^{-2}$
s$^{-1}$ \AA$^{-1}$. The one on the right is normalized by the mean flux in the wavelength
interval displayed. Both spectra are displayed on the same normalized flux scale.  
The wavelength scale $\lambda_0$ shows vacuum wavelength in the rest frame. The dotted line
represents the off-nuclear spectrum broadened to match the adopted velocity dispersion at the 
nucleus.
This spectrum is shifted downward by 1.6 normalized flux units. 
(b) The spectrum of the nuclear point source from (a) corrected for Galactic and
internal extinction. A fit to the continuum (a sixth order polynomial) is shown as a dashed line. 
\label{fig4}}

\figcaption{(a) The input spectra from which the synthetic spectra are constructed.
Both the A0~V and power-law nonstellar spectra  
are normalized to unit flux at 5556~\AA. (b)
The synthetic spectra constructed from a linear combination of the input spectra in (a)
with Gaussian noise added. Each spectrum is labeled with the fraction of the light originating
from A stars. 
The flux scale on the left applies to the spectrum where
10\% of the light originates from A stars. The others are shifted vertically for clarity. 
The fit to the
continuum of the observed spectrum (Fig.~4b) is superimposed for comparison. \label{fig5}}

\figcaption{The emission line ratios of M84's nucleus. The closed circles represent the
line fluxes from Tables 1 and 2 corrected for both Galactic and internal reddening. For line ratios 
involving H$\beta$, 
the open circle would be the position
of M84 if the correction for H$\beta$ absorption is equal to the upper limit derived in the
text. The true position of M84 must be between these two points. For 
the [O~II] $\lambda 3727$/[O~III] $\lambda 5007$ plot, the flux ratio from Table 1 corrected only 
for
Galactic extinction is represented by the open diamond. M87 is represented by closed
and open triangles. The closed triangle
shows the average emission line ratios in the central $0\farcs78$ = 64 pc (Chakrabarti~1995, from
the spectra of Harms et al.~1994), 
while the open triangle
represents the emission line ratios at $0\farcs6$ (49 pc) from the nucleus along the major axis
of its gas disk (Dopita et al.~1997). The error bars claimed for M87 are not much larger than the 
size of
the points. \label{fig6}} 

\figcaption{The spectral energy distribution (SED) 
of M84's nuclear point source (filled points and upper limits), including our nuclear spectrum 
(short
solid line)
from Fig.~4b after it was convolved with a median filter 11 pixels wide (to emphasize the continuum
shape). The dotted line represents the mean SED of
radio loud quasars (Elvis et al.~1994). The small open circles represent the SED of BL Lac (Bregman 
et al.~1990); these points are connected by the dashed line. Both the radio loud quasar and BL Lac
SEDs have been normalized to M84's SED at K band. We use the dashed line to estimate the bolometric
luminosity of M84's nuclear point source in the text. \label{fig7}}

\clearpage

\begin{deluxetable}{lcccccc}
\footnotesize
\tablewidth{468pt}
\tablecaption{Emission Line Fluxes for the Nuclear Spectrum \label{tbl-1}}
\tablehead{
\colhead{Emission Line}  & \colhead{$F_{Cont}^G$} & \colhead{$F^G$} &
\colhead{$F_{Cont}^{G,i}$} & \colhead{$F^{G,i}$} & \colhead{EW} &
\colhead{FWHM} \cr
\colhead{} & \colhead{} & 
\colhead{} & \colhead{} & \colhead{} & \colhead {(\AA)} &
\colhead{(km s$^{-1}$)}} 
\startdata
[O II] $\lambda 3727$ & $0.37 \pm 0.09$ & $2.93 \pm 0.33$ & 
$1.20 \pm 0.32$ & $9.44 \pm 1.08$ & $78 \pm 8$ & 1325 \nl
[Ne III] $\lambda 3869$ & $0.44 \pm 0.08$ & $0.24 \pm 0.18$ &
$1.37 \pm 0.27$ & $0.76 \pm 0.57$ & $5 \pm 4$ & 756 \nl
[S II] $\lambda\lambda 4069,4076$ & $0.69 \pm 0.04$ & $1.05 \pm 0.33$ &
$2.08 \pm 0.13$ & $3.15 \pm 0.99$ & $15 \pm 4$ & 2625 \nl
H$\beta$ $\lambda 4861$ & $1.10 \pm 0.03$ & $0.85 \pm 0.14$ &
$2.73 \pm 0.08$ & $2.11 \pm 0.35$ & $7 \pm 1$ & 1299 \nl
[O III] $\lambda 4959$ & $1.12 \pm 0.02$ & $1.14 \pm 0.17$ &
$2.73 \pm 0.07$ & $2.78 \pm 0.42$ & $10 \pm 1$ & \nodata \nl
[O III] $\lambda 5007$ & $1.14 \pm 0.02$ & $3.48 \pm 0.17$ &
$2.73 \pm 0.07$ & $8.31 \pm 0.42$ & $30 \pm 1$ & 1662 \nl  
\enddata
\tablecomments{$F_{Cont}^G$ and $F_{Cont}^{G,i}$ are the continuum fluxes in units of
$10^{-16}$ erg cm$^{-2}$ s$^{-1}$ \AA$^{-1}$ corrected respectively for Galactic extinction,
and Galactic plus internal extinction. $F^G$ and $F^{G,i}$ are the emission line
fluxes in units of $10^{-15}$ erg cm$^{-2}$ s$^{-1}$ corrected respectively for
Galactic extinction, and Galactic plus internal extinction. EW is the equivalent 
width.}
\end{deluxetable}

\clearpage

\begin{deluxetable}{lcc}
\footnotesize
\tablewidth{250pt}
\tablecaption{Relative Emission Line Fluxes \label{tbl-2}}
\tablehead{
\colhead{Emission Line}  & \colhead{$F^G$} & \colhead{$F^{G,i}$}} 
\startdata
[O I] $\lambda 6300$ & 0.12 & 0.13 \nl
[N II] $\lambda 6548$ & 0.42 & 0.43 \nl
H$\alpha$ & 1.00 & 1.00 \nl
[N II] $\lambda 6583$ & 1.29 & 1.29 \nl
[S II] $\lambda 6717$ & 0.51 & 0.50 \nl
[S II] $\lambda 6731$ & 0.58 & 0.57 \nl  
\enddata
\tablecomments{$F^G$ and $F^{G,i}$ are the relative fluxes corrected
respectively for Galactic extinction, and Galactic plus internal extinction.}
\end{deluxetable}

\clearpage

\begin{deluxetable}{lcccccc}
\footnotesize
\tablewidth{400pt}
\tablecaption{HST V Band Magnitudes for M84's Nuclear Point Source \label{tbl-3}}
\tablehead{
\colhead{Date}  & \colhead{HST mode} & \colhead{Spectral Element} &
\colhead{$\lambda_{\rm eff}$} & \colhead{$\Delta\lambda$} &
\colhead{V$^a$} & \colhead{Reference} \cr
& & & \colhead{(\AA)} & \colhead{(\AA)} & \colhead{(mag)} } 
\startdata
06 Mar 1993 & WFPC & F555W & 5454 & 509 & 20.2 & 1 \nl
04 Mar 1996 & WFPC2 & F547M & 5454 & 486 & 19.9 & 2 \nl
10 Apr 1998 & STIS/ACCUM & G430L & 5454 & 497 & 19.6 & 3 \nl 
\enddata
\tablenotetext{a}{Corrected for Galactic extinction but not internal extinction.}
\tablerefs{(1) Jaffe et al.~1994. (2) Paper I.
(3) This paper.}
\end{deluxetable}

\clearpage

\begin{deluxetable}{lcc}
\footnotesize
\tablewidth{330pt}
\tablecaption{The SED of M84's Nuclear Point Source \label{tbl-4}}
\tablehead{
\colhead{$\nu$}  & \colhead{F$_{\nu}^a$} & \colhead{Reference} \cr
\colhead{(Hz)} & \colhead{(erg cm$^{-2}$ s$^{-1}$ Hz$^{-1}$)} & } 
\startdata
$1.67 \times 10^9$ & $1.60 \times 10^{-24}$ & 1 \nl
$8.09 \times 10^9$ & $1.90 \times 10^{-24}$ & 2 \nl
$2.00 \times 10^{13}$ & $< 5.00 \times 10^{-26}$ & 3 \nl
$3.11 \times 10^{13}$ & $< 6.08 \times 10^{-26}$ & 3 \nl
$4.44 \times 10^{13}$ & $< 1.02 \times 10^{-25}$ & 3 \nl
$1.46 \times 10^{14}$ & $4.66 \times 10^{-27}$ & 3 \nl
$1.87 \times 10^{14}$ & $4.76 \times 10^{-27}$ & 3 \nl
$2.72 \times 10^{14}$ & $2.89 \times 10^{-27}$ & 3 \nl
$3.63 \times 10^{14}$ & $1.36 \times 10^{-27}$ & 4 \nl
$5.50 \times 10^{14}$ & $8.83 \times 10^{-28}$ & 4 \nl
$1.30 \times 10^{15}$ & $< 9.55 \times 10^{-28}$ & 5 \nl
$4.84 \times 10^{17}$ & $< 1.84 \times 10^{-30}$ & 6 \nl
\enddata
\tablenotetext{a}{Corrected for Galactic and internal extinction 
(total $A_V = 0.94$).}
\tablerefs{(1) Jones et al.~1981. (2) Schilizzi 1976.
(3) This paper. (4) Paper I. (5) Zirbel \& Baum 1998.
(6) Fabbiano et al.~1992.}
\end{deluxetable}

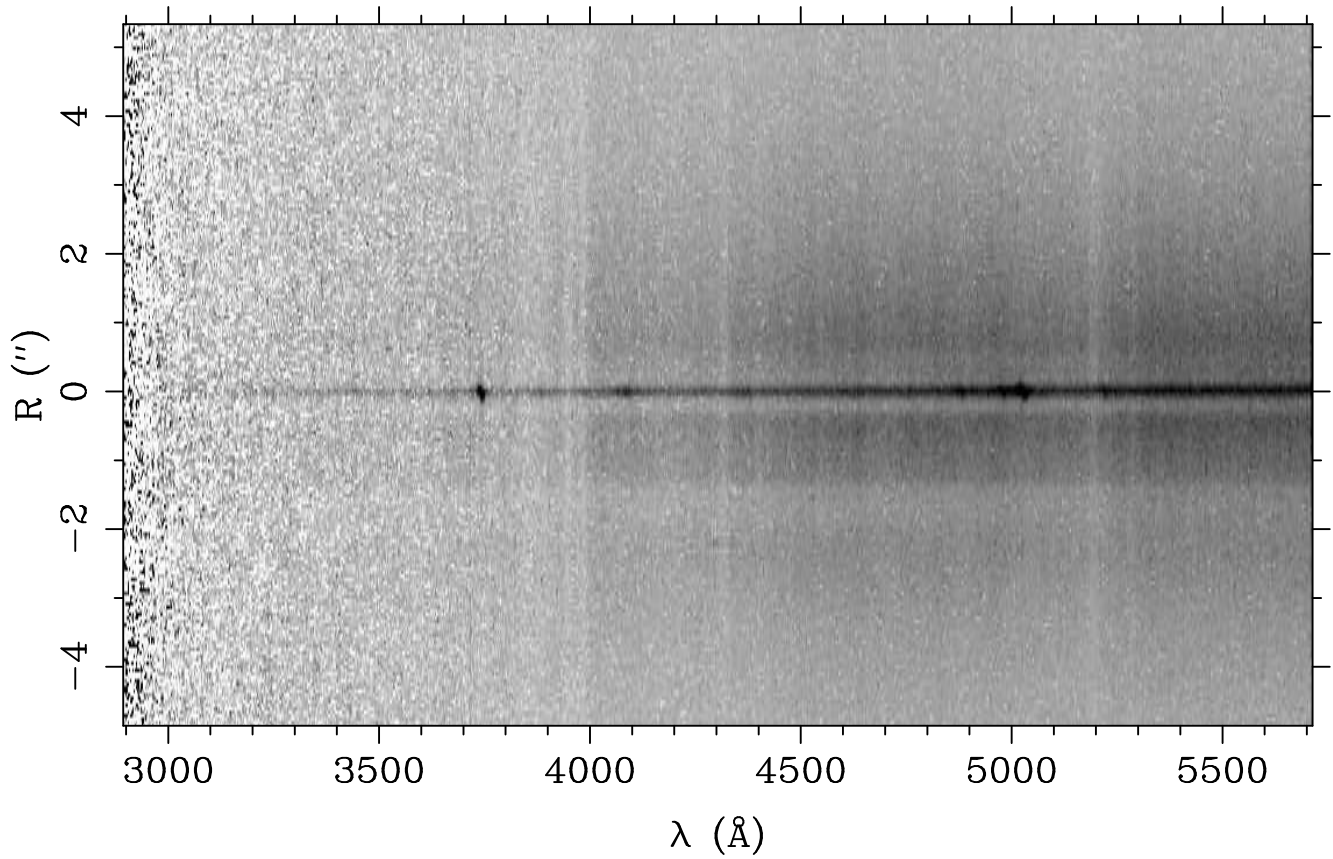
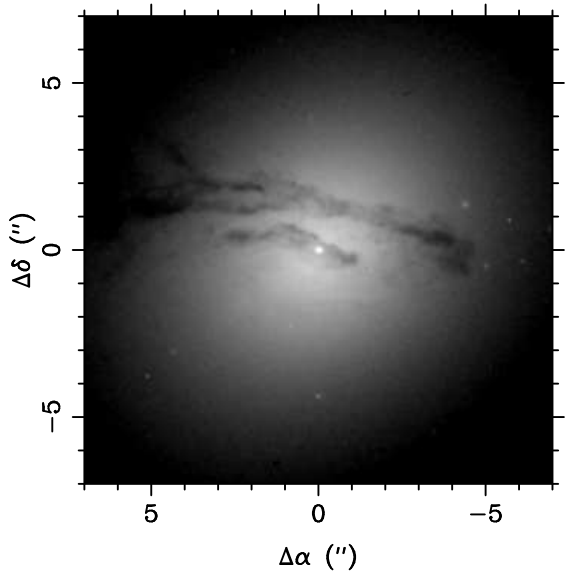
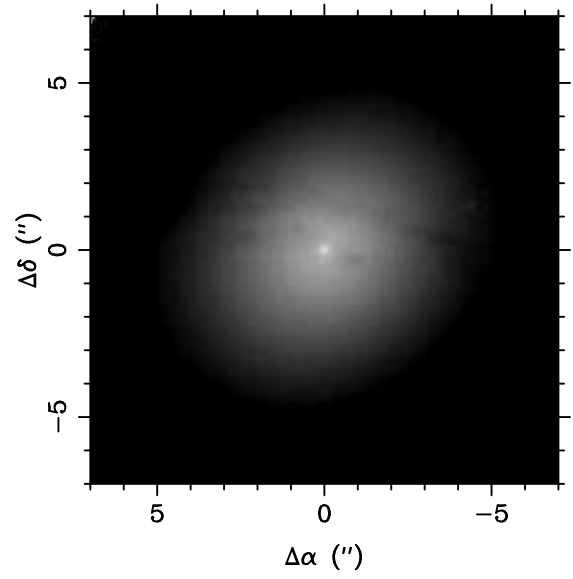
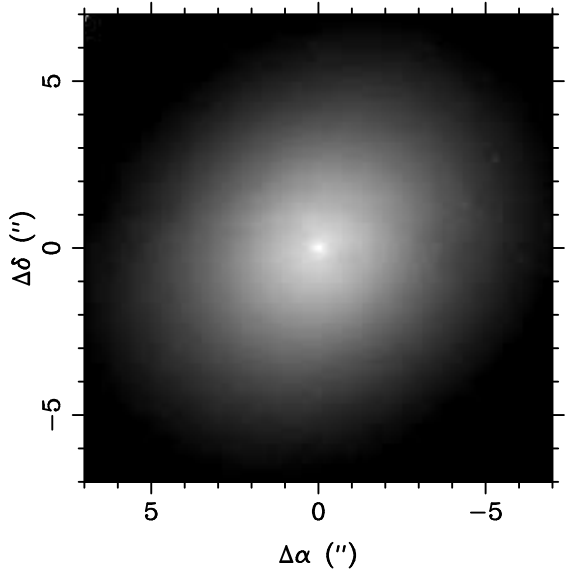
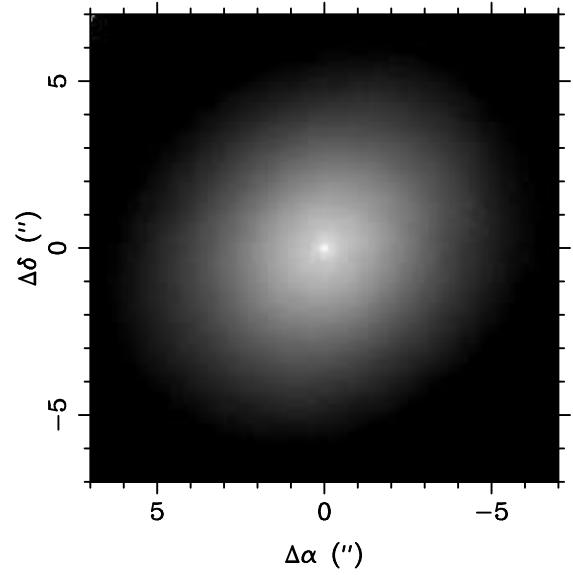
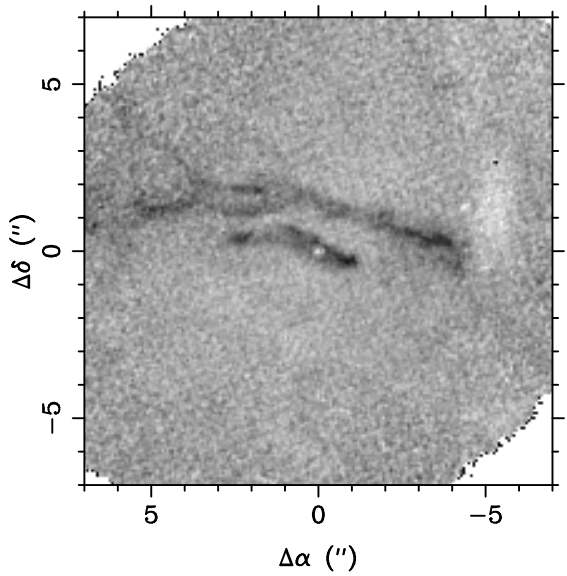
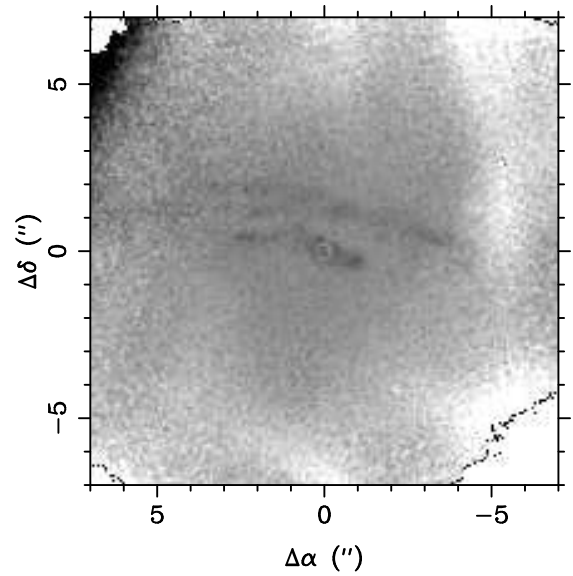
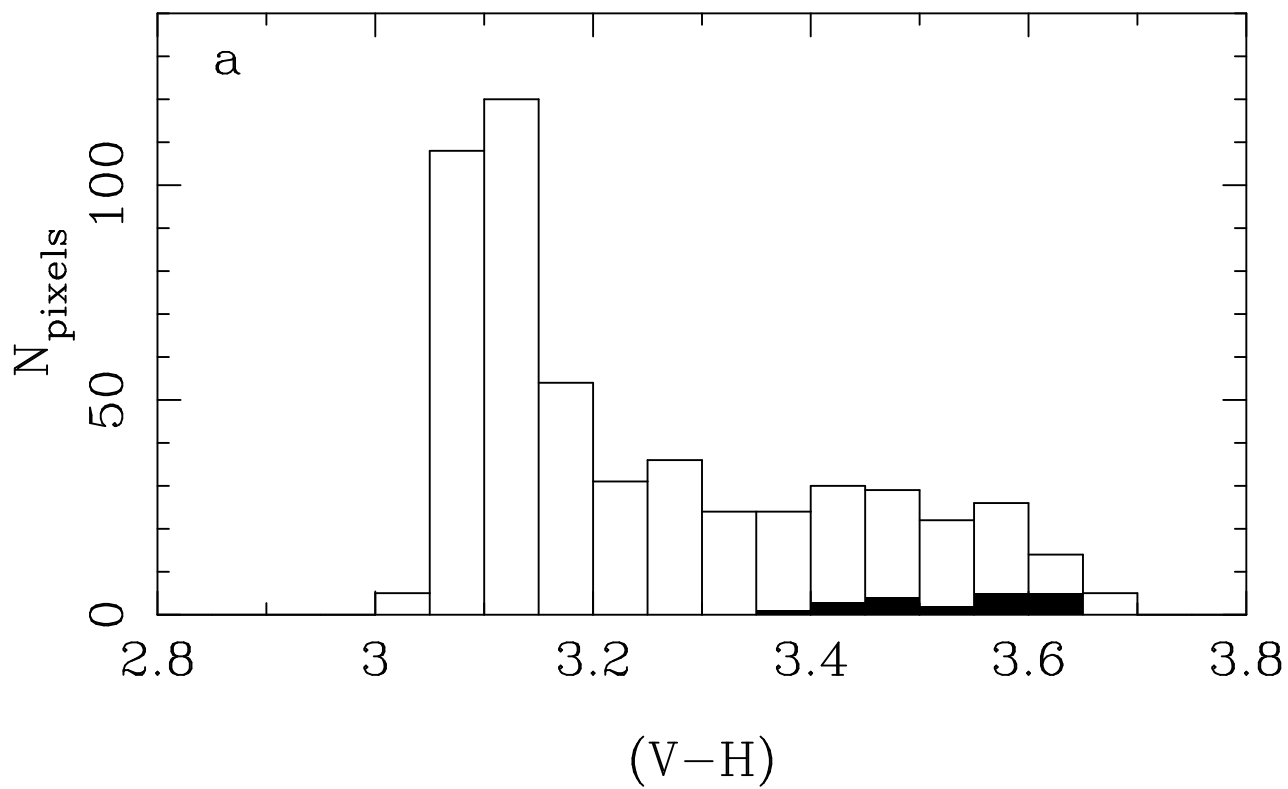
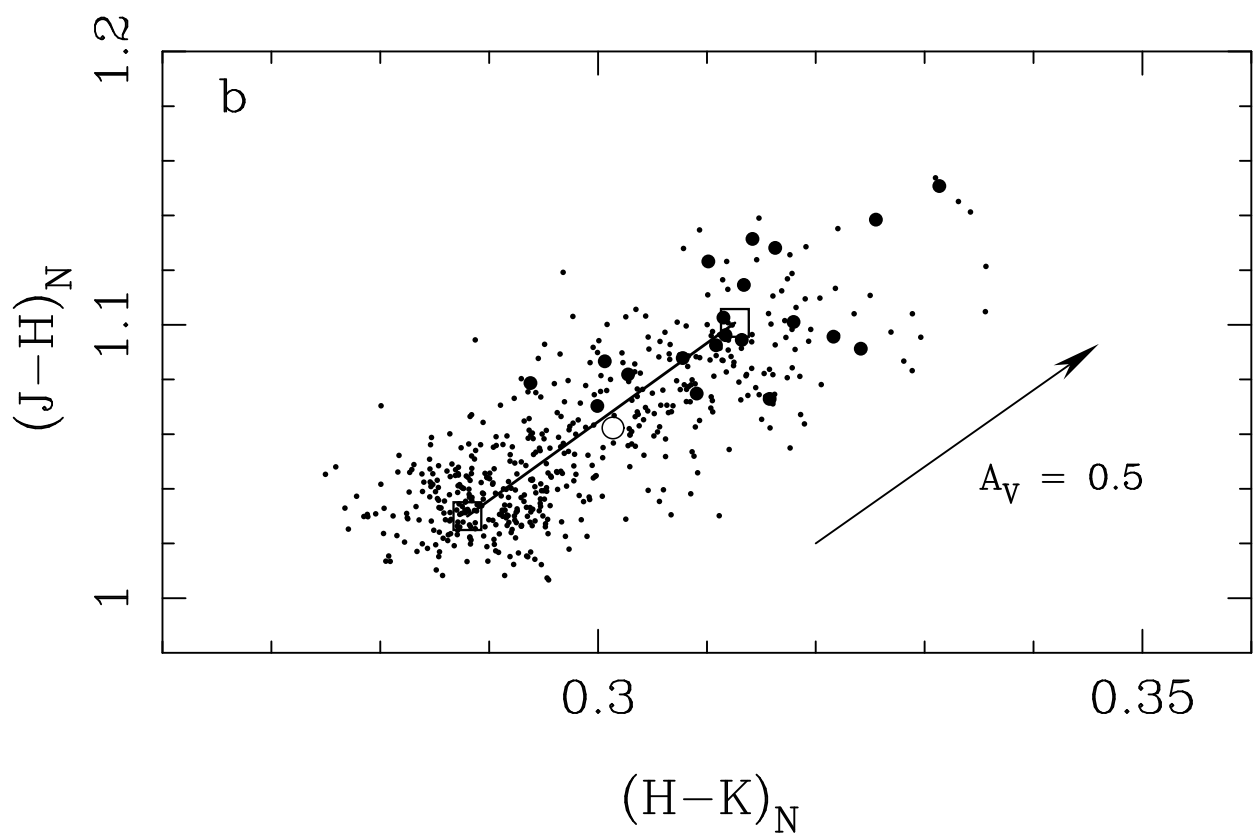
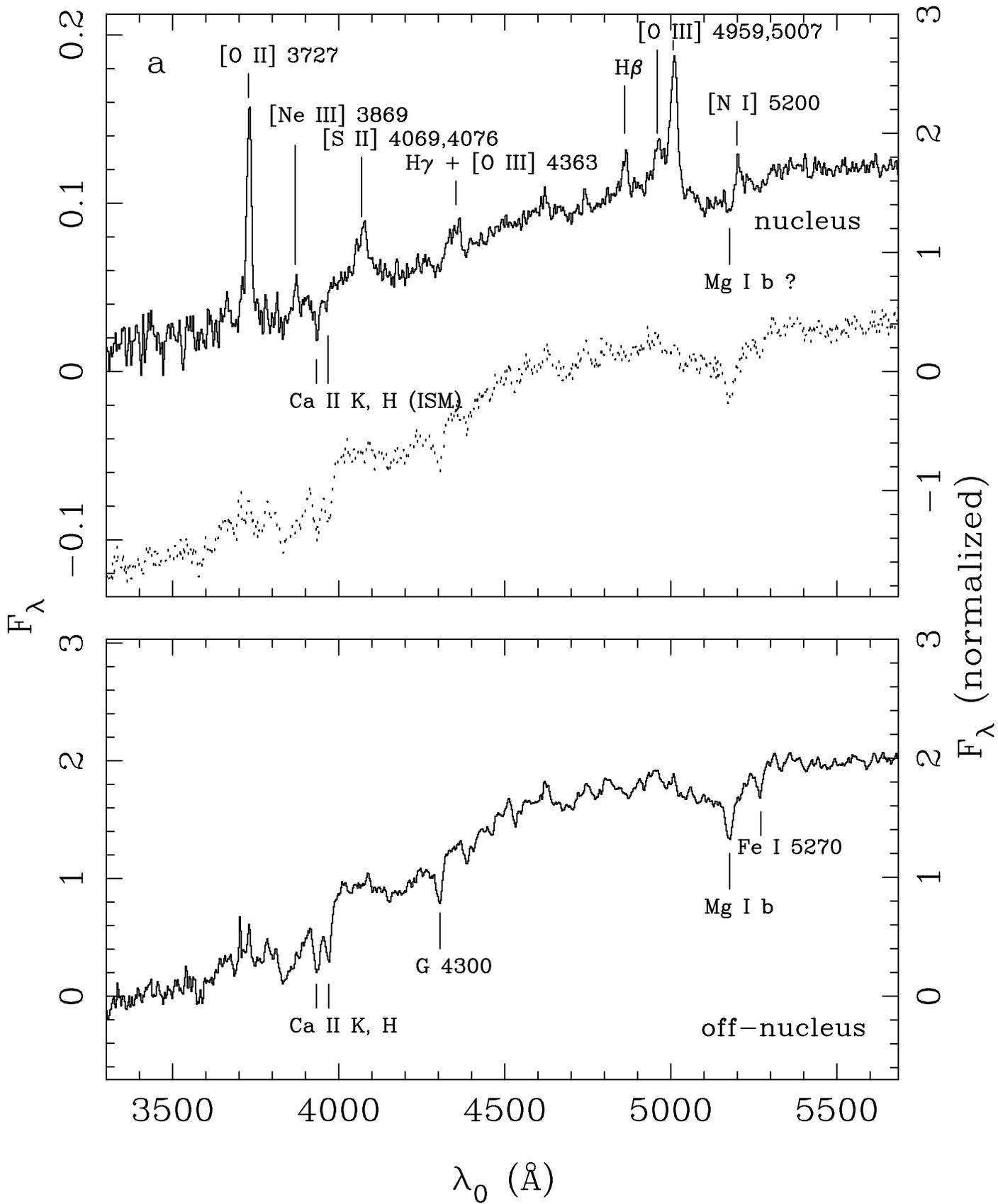
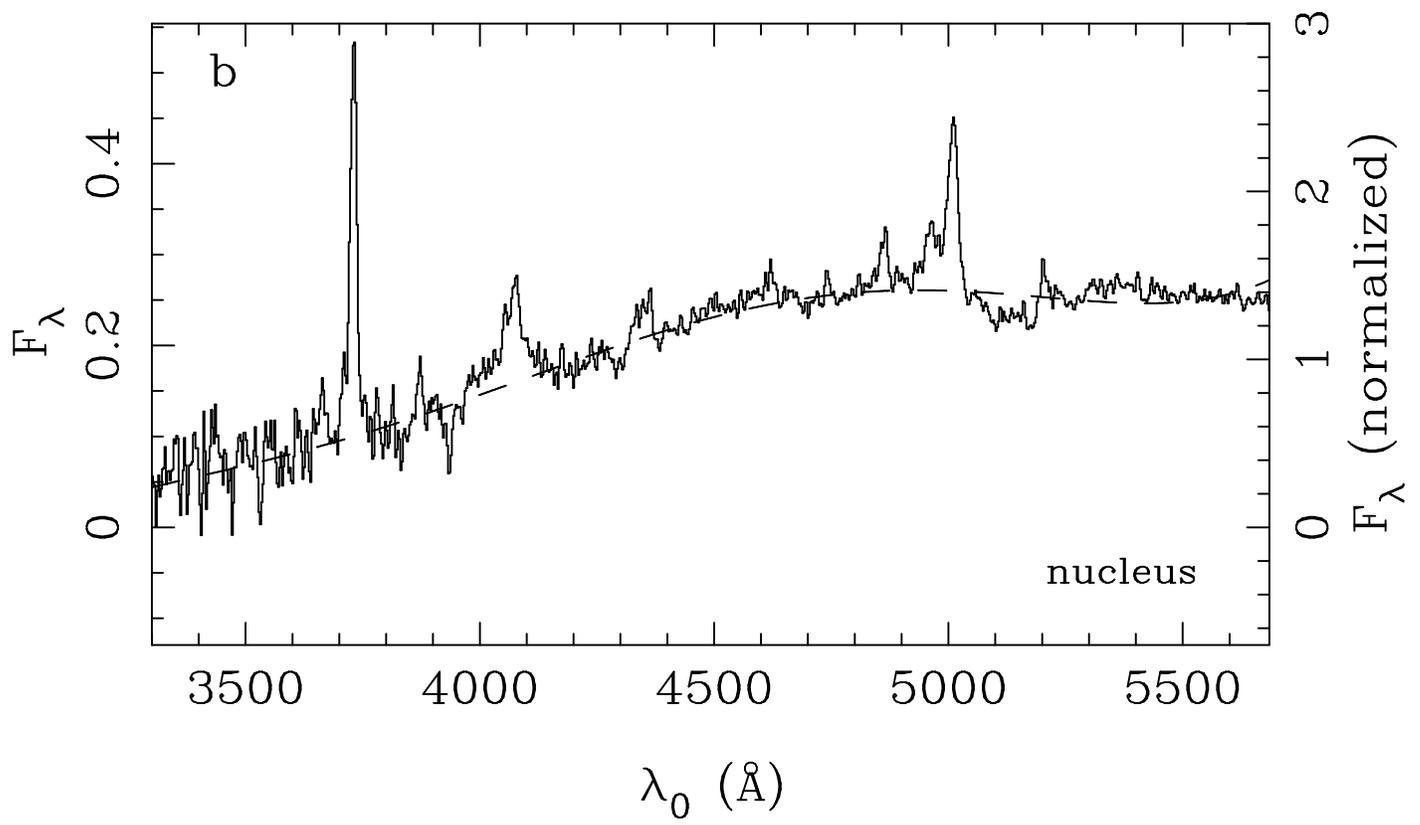
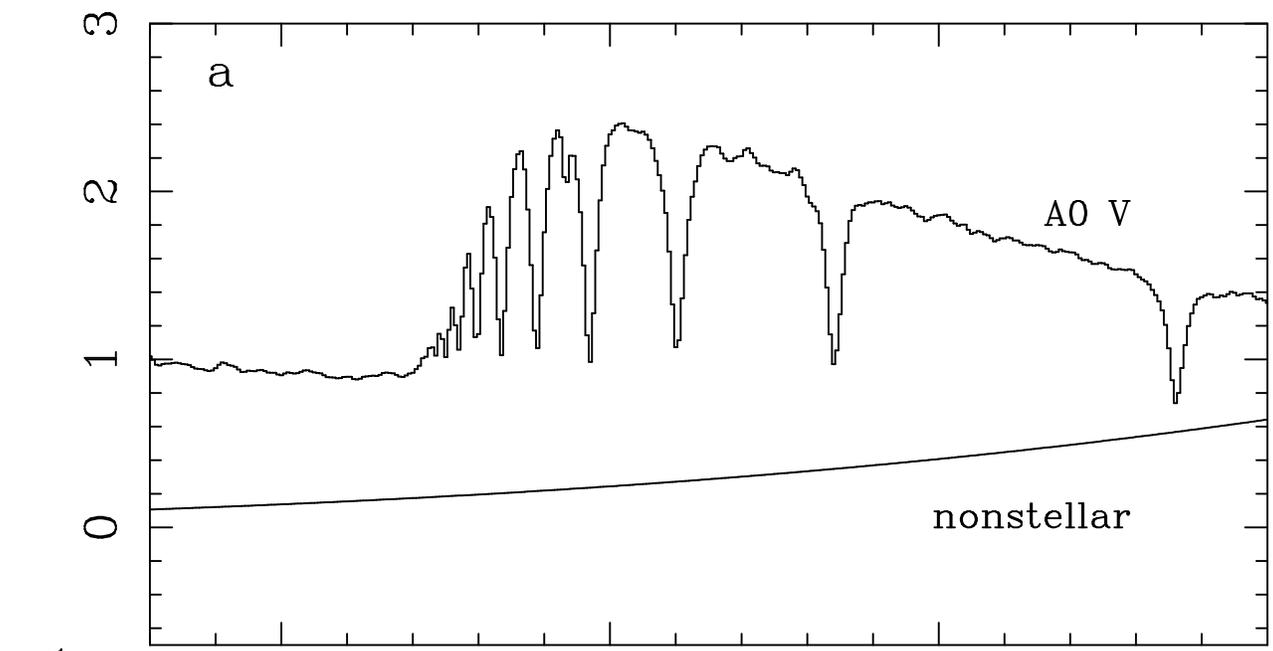
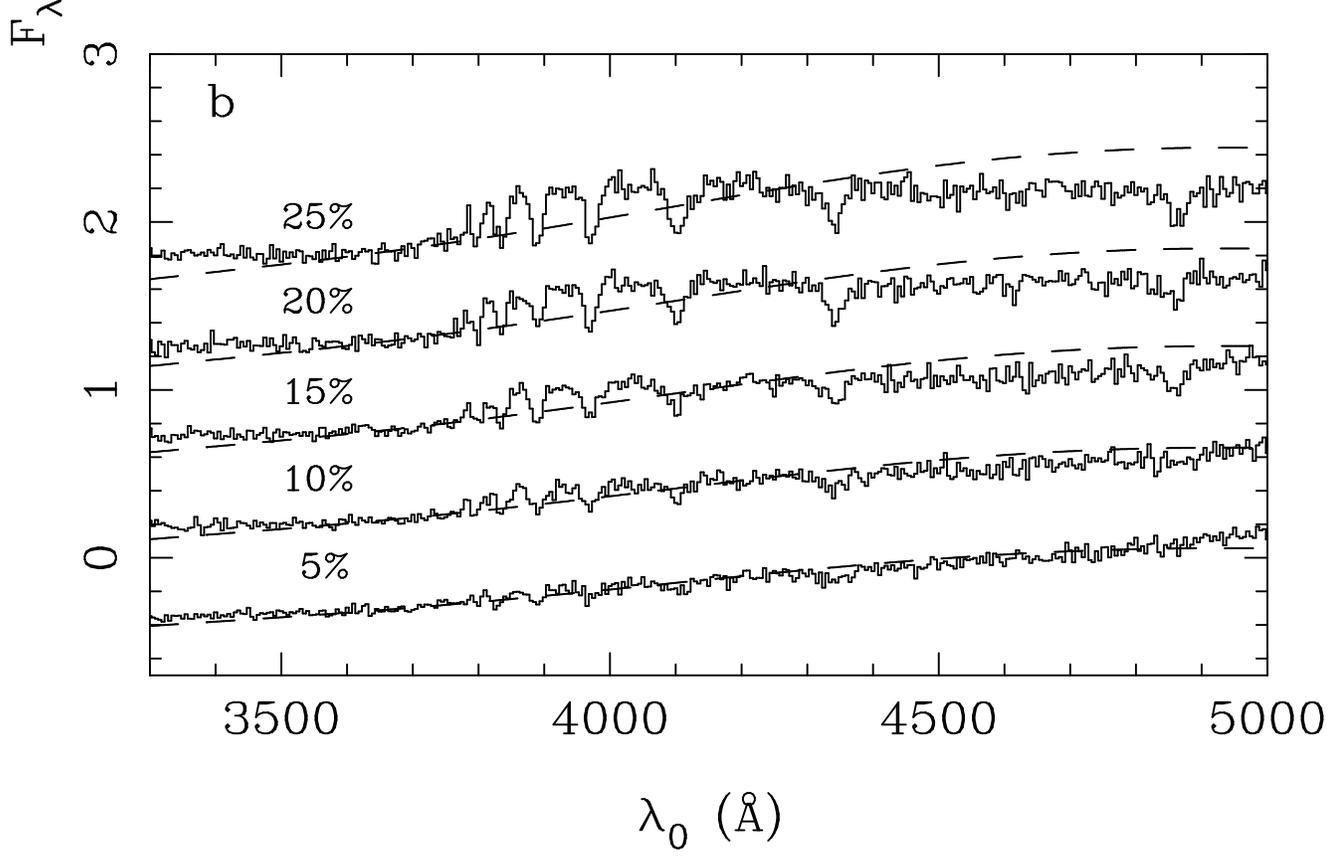
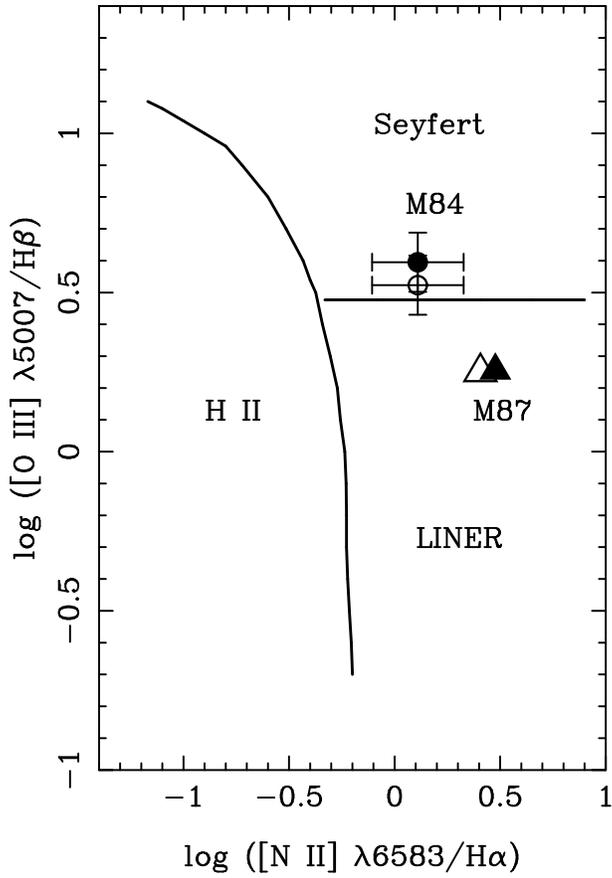
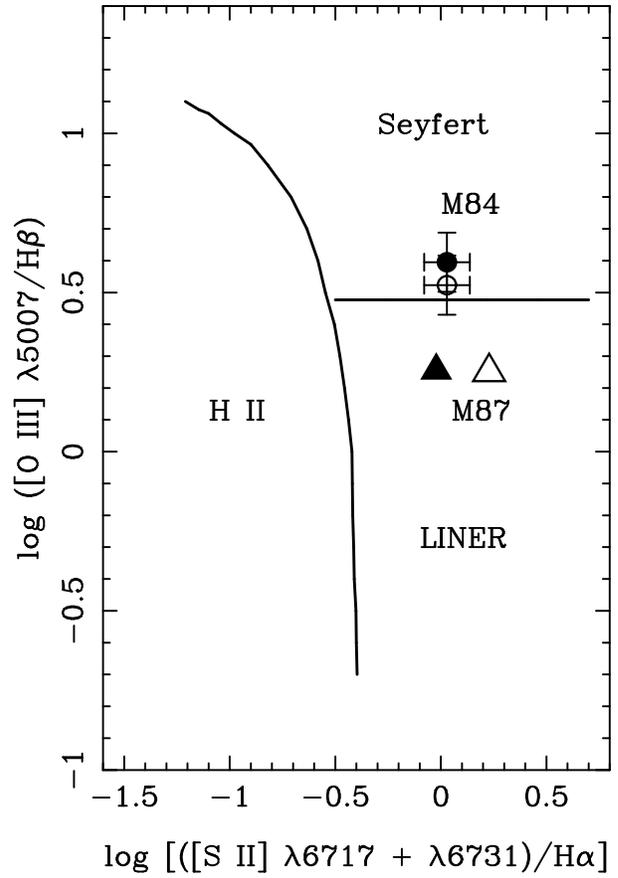
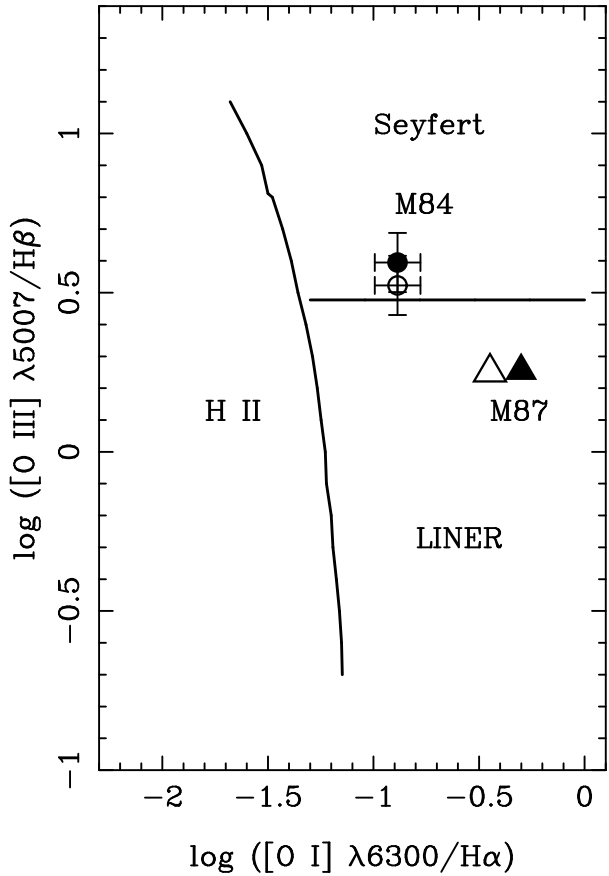
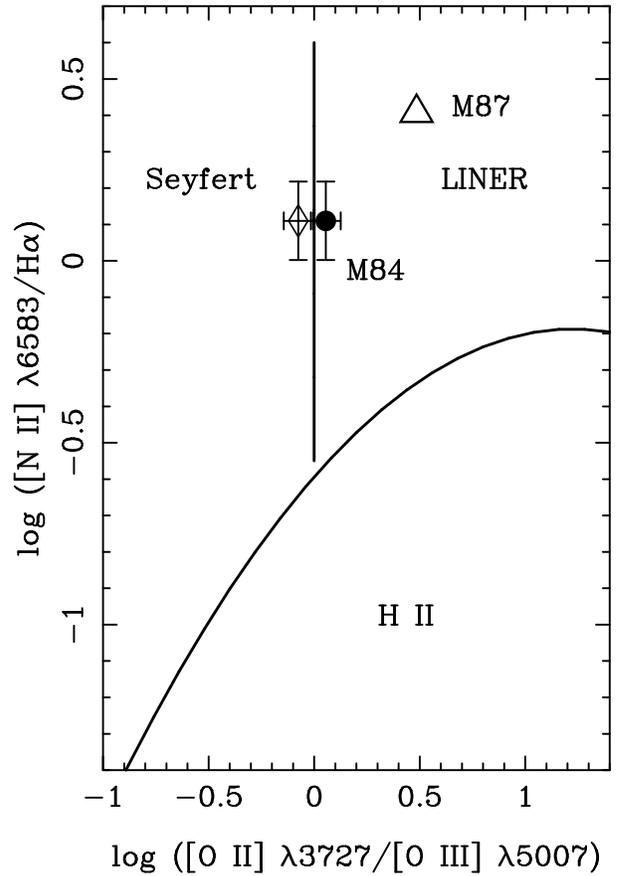
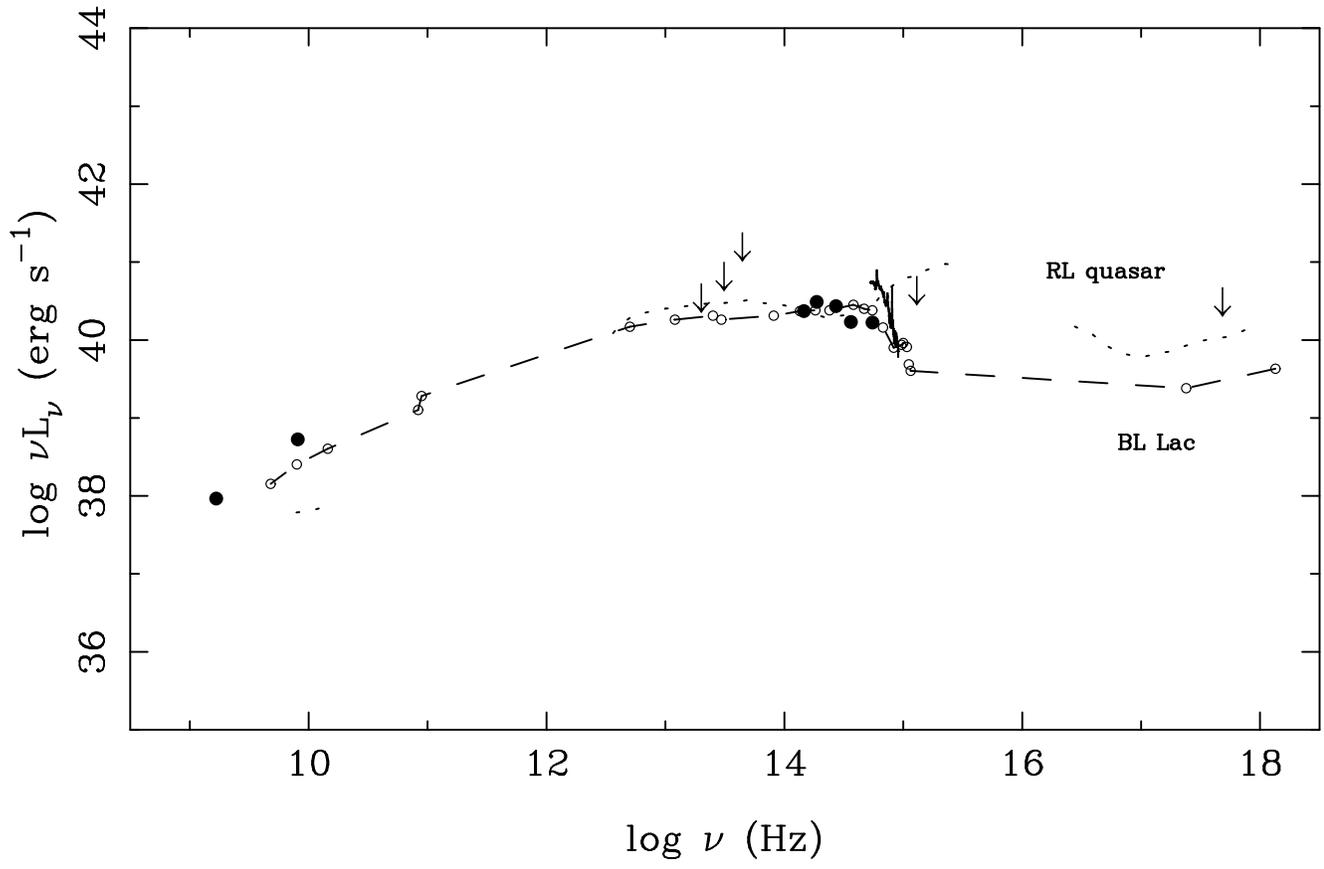
\end{document}